\documentclass[prb,twocolumn,tightenlines,citeautoscript,%
showpacs,floatfix,superscriptaddress]{revtex4}

\usepackage{bm}
\usepackage{amssymb}
\usepackage{graphicx}
\usepackage{amsmath}
\usepackage{textcomp}
\usepackage{lscape}

\begin{document}
\title{Theory of optical spin orientation in silicon}
\author{J. L. Cheng}
\affiliation{Department of Physics and Institute for Optical Sciences,
University of Toronto, 60 St. George Street, Toronto, Ontario, Canada
M5S 1A7}
\affiliation{Hefei National Laboratory for Physical Sciences at
    Microscale, University of Science and Technology of China, Hefei,
    Anhui, 230026, China}
\author{J. Rioux}
\affiliation{Department of Physics and Institute for Optical Sciences,
University of Toronto, 60 St. George Street, Toronto, Ontario, Canada
M5S 1A7}
\author{J. Fabian}
\affiliation{Institute for Theoretical Physics, University of
  Regensburg, 93040 Regensburg, Germany}
\author{J. E. Sipe}
\thanks{Author to whom correspondence should be addressed}%
\email{sipe@physics.utoronto.ca}
\affiliation{Department of Physics and Institute for Optical Sciences,
University of Toronto, 60 St. George Street, Toronto, Ontario, Canada
M5S 1A7}

\date{\today}
\begin{abstract}
  We theoretically investigate the indirect optical injection of carriers and
  spins in bulk silicon, using an empirical pseudopotential
  description of electron states and an adiabatic bond charge model for
  phonon states. We identify
  the selection rules, the contribution to the carrier and spin
  injection in each conduction band valley from each phonon branch and
  each valence band, and the temperature dependence of
  these processes. The transition from the heavy hole band to the
  lowest conduction band dominates the injection due to 
  the large joint density of states. For incident light propagating along
  the $[00\bar{1}]$ direction, the injection rates and the degree of
  spin polarization of injected electrons show strong valley
  anisotropy. The maximum degree of spin polarization is at the
  injection edge with values $25\%$ at low temperature and $15\%$ at
  high temperature. 
\end{abstract}


\maketitle

\section{Introduction}
The optical injection of carriers is a powerful method for
the study of the properties of semiconductors, and the optical injection of
spins, {\it i.e.}, optical orientation\cite{optical_orientation}, is
an important element in the toolkit of the field of spintronics
\cite{Rev.Mod.Phys._76_323_2004_Zutic,ActaPhys.Slovaca_57_565_2007_Fabian,Phys.Rep._493_61_2010_Wu}. Most
research has been focused on direct optical 
transitions. There have been fewer studies of indirect optical
transitions,in which the excited electron and hole have
  different wavevectors,
and phonon emission or absorption processes are necessary to conserve
the total wave vector. In this paper, we consider the indirect optical
 injection of carriers and spins in bulk silicon. 

Macfarlane {\it et al.}\cite{Phys.Rev._111_1245_1958_Macfarlane} first
measured the fine structure of the absorption-edge spectrum in
intrinsic bulk germanium and silicon in 1958, and identified the phonon-assisted indirect gap absorption
branches. Thereafter, these studies were extended to doped
silicon\cite{Prog.Photovolt:Res.Appl._3_189_1995_Green,Phys.Rev.Lett._70_3659_1993_Suemoto,Appl.Phys.Lett._93_131916_2008_Mendeleyev,J.Appl.Phys._61_4923_1987_Saritas,J.Appl.Phys._50_1491_1979_Weakliem}.
Theoretically, Elliott\cite{Phys.Rev._108_1384_1957_Elliott} was the
first to study indirect absorption of excitons under the effective mass
approximation, and identified the absorption lineshapes at
  photon energy near and far away from the indirect
  gap. He found that the lineshapes are not sensitive to exciton effects at
  high photon energy. The electroabsorption in indirect gap
  semiconductor\cite{Phys.Rev.B_4_4424_1971_Lao,Phys.Rev.Lett._26_499_1971_Lao}
  and the absorption spectra of multiexciton-impurity complexes
  \cite{Sov.Phys.Usp._24_815_1981_Kulakovskii} were also investigated.
Hartman\cite{Phys.Rev._127_765_1962_Hartman} determined the absorption spectra using
a parabolic band approximation. Dunn\cite{Phys.Rev._166_822_1968_Dunn}
and
Chow\cite{Phys.Rev._185_1056_1969_Chow,Phys.Rev._185_1062_1969_Chow}
used a Green function method to investigate the physical processes in
indirect absorption. All these models approximated the matrix
elements of the electron-phonon interaction by their band edge values. Later, pseudopotential models
\cite{Phys.Rev.Lett._48_413_1982_Glembocki,Phys.Rev.Lett._48_1296_1982_Bednarek,Phys.Rev.B_42_5714_1990_Li,Phys.Rev.B_25_1193_1982_Glembocki,AnnalenderPhysik_504_24_1992_Klenner}
were employed to calculate the transition matrix elements of the electron-phonon
interaction around the band edge. However, a full band structure calculation of the 
  full spectrum of indirect gap absorption is still absent, even with
  the neglect of the excitonic effects. 

Investigations of indirect optical spin injection are less common than
those of indirect gap carrier injection. They are also less common
than those of direct gap spin injection\cite{Phys.Rev.B_76_205113_2007_Nastos}, despite the fact that the first
optical orientation experiment\cite{Phys.Rev.Lett._20_491_1968_Lampel}, performed 
by Lampel in his study of the nuclear polarization of $^{29}$Si in bulk
silicon, employed indirect absorption. The degree of spin polarization
in such an injection process is sometimes understood as a spin-dependent virtual optical
transition combined with a spin-independent phonon absorption or
emission process\cite{Proc.9thInt.Coef.Phys.Semicond._2_1139_1968_Lampel,PhdThesis_Verhulst_2004}.
However, due to spin-orbit coupling the electron states are not
pure spin eigenstates, and the effect of the electron-phonon
interaction on the indirect gap injection needs to be calculated in detail. 
While Li and Dery \cite{Phys.Rev.Lett._105_037204_2010_Li} have
recently studied the degree of circular polarization of the
luminescence associated with the recombination across the indirect band gap following the electrical
injection of spins in silicon, a theoretical investigation of indirect optical
spin injection is still absent. 

In the present paper, we perform a full band structure calculation of the
indirect optical injection of carriers and spins using an empirical
pseudopotential model
\cite{Phys.Rev.B_10_5095_1974_Chelikowsky,Phys.Rev.B_14_556_1976_Chelikowsky,Phys.Rev._149_504_1966_Weisz}
(EPM) for electron states and an adiabatic bond charge model
\cite{Phys.Rev.B_15_4789_1977_Weber} (ABCM) for
phonon states. Compared to $\bm k\cdot\bm p$ models and {\it ab initio}
models, which are widely used for direct gap carrier and spin
injection calculation,\cite{Phys.Rev.B_76_205113_2007_Nastos,Phys.Rev.B_81_155215_2010_Rioux}
the advantage of the EPM is that the electron-phonon interaction in the
whole Brillouin zone can be calculated consistently in combination
with the ABCM. This approach has been successfully used to describe
spin relaxation processes \cite{Phys.Rev.Lett._104_016601_2010_Cheng} and 
photoluminescence\cite{Phys.Rev.Lett._105_037204_2010_Li} in bulk silicon.

We focus on the calculated optical indirect injection coefficients of
carriers and spins, identifying the contribution from each valence band and
phonon branch. We take the excited electrons and holes as
free carriers. Because the electron-hole interaction is nearly
spin-independent, to good approximation it should affect the spin
injection and the carrier injection rates in the same way, and not
affect the degree of spin polarization (DSP) of the injected
electrons, which is the ratio of these two quantities. The dependence of the injection coefficients and DSP on photon
energy, conduction band valley, and temperature are established. For the injection of carriers,
our numerical results agree with the experiments at high
photon energy. In the course of our investigations we also discuss in
detail a simple but widely used model, in which the values of the
transition matrix elements are approximated by their values at the
band edge, and we compare its predictions with our calculations.

We organize the paper as follow: We begin with the general formula for indirect
carrier and spin injection assisted by phonon emission and absorption in
bulk silicon in Sec.~\ref{sec:model}. Then the selection rules that follow from the crystal symmetry
are discussed in Sec.~\ref{sec:selectionrules}. We introduce the EPM and ABCM
models in Sec.~\ref{sec:epm}. Finally we present our results and
conclusions in Sec.~\ref{sec:cal} and Sec.~\ref{sec:result}. 
In an Appendix, we describe an improved adaptive linear analytic
tetrahedral integration method
(LATM)\cite{Phys.Rev.B_49_16223_1994_Blochl,Phys.Rev.B_76_205113_2007_Nastos}
that we use to perform the six-fold integration over the Brillouin zone (BZ). 

\section{Model For  carrier and spin injection by indirect absorption}
\label{sec:model}
\begin{figure}[htp]
  \centering
  \includegraphics[height=6cm]{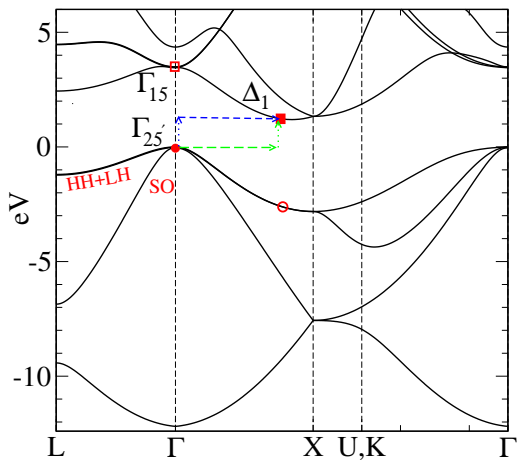}
  \caption{(color online). Band structure of silicon, calculated from
    the EPM, with symmetry notation at the $\Gamma$ point and the
    conduction band edge ($\Delta_1$). The valence band 
    is noted by HH, LH, and SO bands. Two important phonon-assisted
    transitions from valence band edge (red dot) to conduction band
    edge (red filled square) are given by green lines and blue lines. The
    dotted lines refer to optical transitions, while the dashed
    lines refer to the transitions mediated by phonons. The red hollow
    square (circle) stands for one of the possible intermediated states in the
    transition given by blue (green) lines. }
\label{fig:bandstructure}
\end{figure}
Fig.~\ref{fig:bandstructure} shows the band structure of silicon in
the energy range $[-12~\text{eV}, 6~\text{eV}]$ calculated from the
EPM; the details are given in section \ref{sec:epm}. 
The lowest four bands are valence bands, of which the upper three
are the heavy hole (HH), light hole (LH), and spin split-off (SO) bands. 
The valence band edge (red dot) is at the $\Gamma$ point with $\bm k_v^0=\bm 0$. All other bands are
conduction bands. From the figure, the conduction band edge (red
filled square) is at $\bm k_c^0\approx0.85\overrightarrow{\Gamma
  X}$ on the $\Delta$ symmetry line, and results in six equivalent
valleys that can be denoted as $X,\bar{X},Y,\bar{Y},Z,\bar{Z}$,
indicating the location of the valley center. The calculated direct band gap
at the $\Gamma$ point is $E_g=3.4$~eV,  while the indirect
band gap is $E_{ig}=1.17$~eV. When the photon energy satisfies
$E_{ig}<\hbar\omega<E_{g}$, optical injection
occurs only across the indirect gap. Because the excited electron and hole have
different wave vectors, the transition must be assisted by phonon
emission or absorption. The green and blue lines show two possible indirect gap transitions
between the valence and conduction band edges. The red hollow square
and circle indicate possible intermediate states.

For an electric field ${\bm E}(t) = {\bm
  E}_\omega e^{-i\omega t} +  c.c$, the carrier and spin injection
rates in silicon can be generally written as  
\begin{eqnarray}
  \dot{n}(T, \omega) &=& \xi^{ab}(T, \omega)E_\omega^a
  \left(E_\omega^b\right)^{\ast}\ ,\nonumber\\
  \dot{S}^f(T, \omega) &=& \zeta^{fab}(T,
  \omega)E_\omega^a\left(E_\omega^b\right)^{\ast}\ .
\label{eq:injection}
\end{eqnarray}
Here $\xi^{ab}(T, \omega)$ and $\zeta^{fab}(T, \omega)$ are the injection coefficients for carriers and
spins, respectively,  at temperature $T$. The superscript Roman characters indicate Cartesian coordinates,
and repeated superscripts are to be summed over. 

Using Fermi's Golden Rule, we find the injection coefficient  $\xi^{ab}$ and
$\zeta^{fab}$ to be of the form 
\begin{equation}
{\cal
  A}^{ab}=\sum_{I;cv\lambda\pm}{\cal A}^{ab}_{I;cv\lambda\pm}\ ,
\label{eq:totala}
\end{equation}
with 
\begin{eqnarray}
  {\cal A}^{ab}_{I;cv\lambda\pm} &=& \frac{2\pi}{\hbar}\sum_{\bm
    k_c\in I, \bm k_v}\delta(\varepsilon_{c\bm
  k_c}-\varepsilon_{v\bm k_v}\pm\hbar\Omega_{(\bm
    k_c-\bm k_v)\lambda}-\hbar\omega) \nonumber\\
  &\times&N_{(\bm k_c-\bm k_v)\lambda\pm} {\cal A}^{ab}_{c\bm k_cv\bm
    k_v;\lambda}\ ,\label{eq:X}
\end{eqnarray}
where
\begin{equation}
{\cal A}^{ab}_{c\bm k_cv\bm k_v;\lambda} =
\sum_{\sigma_c\sigma_c^{\prime}\sigma_v}\langle\bar{c}^{\prime} \bm
k_c|\hat{\cal A}|\bar{c}\bm k_c\rangle T^{a}_{\bar{c}\bm k_c\bar{v}\bm k_v\lambda}\left(T^{b}_{\bar{c}^{\prime}\bm k_c\bar{v}\bm
      k_v\lambda}\right)^{\ast}\ .
\label{eq:transitionmatrix}
\end{equation}
Here $I$ is the valley index; $c$($v$) is the conduction(valence) band
index without including spin;  $\bm k_c$ ($\bm k_v$) is the electron
(hole) wave vector, where $\bm
k_c\in I$ means the summation is over the $I^{th}$ valley;
$\varepsilon_{c\bm k_c}$($\varepsilon_{v\bm k_v}$) give the energy
spectra of conduction (valence) bands; $\hbar\Omega_{\bm q\lambda}$
gives the phonon energy at wave vector $\bm q$ and mode $\lambda$
(longitudinal optical (LO) and acoustic (LA), and transverse optical
(TO) and acoustic (TA) branches); and 
$N_{\bm q\lambda\pm} = N_{\bm q\lambda} + \frac{1}{2}
\pm \frac{1}{2}$, where $N_{\bm q\lambda}$
is the equilibrium phonon number. The operator $\hat{\cal A}$ in
Eq.~(\ref{eq:transitionmatrix}) stands for the identity operator
$\hat{I}$ in the carrier injection calculation, and the $f^{th}$ 
component of the spin operator in the spin injection calculation.
The indirect optical transition matrix elements are
\begin{equation}
  T^{a}_{\bar{c}\bm k_c\bar{v}\bm k_v\lambda} =
  \frac{e}{\hbar\omega}\sum_{\bar{n}}\bigg\{\frac{M_{\bar{c}\bm
      k_c\bar{n}\bm k_v,\lambda}v^a_{\bar{n}\bar{v}\bm k_v}}{\omega-\omega_{nv\bm
      k_v}}+\frac{v^a_{\bar{c}\bar{n}\bm k_c}M_{\bar{n}\bm
      k_c\bar{v}\bm k_v,\lambda}}{\omega_{cn\bm k_c}-\omega}\bigg\} \ ,
\label{eq:T}
\end{equation}
in which $\bar{c}=\{c,\sigma_c\}$,
$\bar{c}^{\prime}=\{c,\sigma_c^{\prime}\}$, $\bar{v}=\{v,\sigma_v\}$,
and $\bar{n}=\{n,\sigma_n\}$ are full band
indexes, with $\sigma_{c},\sigma_c^{\prime},\sigma_v$ being the spin indexes; $\varepsilon_{n\bm k}$ is the electron energy at band $n$ and wave
vector $\bm k$,  and $\omega_{nm\bm k}$ is defined by
$\hbar\omega_{nm\bm k}\equiv\varepsilon_{n\bm 
  k}-\varepsilon_{m\bm k}$. The velocity matrix elements are given by
$\bm v_{\bar{n}\bar{m}\bm k}=\langle \bar{n}\bm k|\hat{\bm
  v}|\bar{m}\bm k\rangle$, with the velocity operator $\hat{\bm v} =
\partial 
  H_e/\partial\bm p$ and the unperturbed electron Hamiltonian $H_e$; $M_{\bar{n}\bm k_c\bar{m}\bm k_v,\lambda}=\langle
\bar{n}\bm k_c|H^{ep}_{\lambda}(\bm k_c-\bm k_v)|\bar{m}\bm k_v\rangle$ are matrix
elements of the electron-phonon interaction $H^{ep}$, which is written as $H^{ep}=\sum_{\bm
  q\lambda}H^{ep}_{\lambda}(\bm q)(a_{\bm q\lambda} + a_{-\bm
  q\lambda}^{\dag})$, with $a_{\bm q\lambda}$ being the
phonon annihilation operator for wavevector $\bm q$ and mode $\lambda$.

We now turn to the symmetry properties of ${\cal
  A}_{I;cv\lambda\pm}^{ab}$. Though bulk silicon has $O_h$ symmetry,
each conduction-band valley only has $C_{4v}$ symmetry\cite{peteryu}.
Therefore each ${\cal A}_{I;cv\lambda\pm}$ only has
$C_{4v}$ symmetry, since the summation over $\bm k_c$ is limited to
$\bm k_c$ in the $I^{th}$ valley. The
summation of Eq. (\ref{eq:X}) can be rewritten as 
\begin{eqnarray}
  {\cal A}_{I;cv\lambda\pm}^{ab} &=& \frac{2\pi}{\hbar}{\cal N}_v{\cal
    N}_{c,I}{\sum_{\bm
    k_c\in I}}^{\prime}{\sum_{\bm k_v}}^{\prime}\frac{1}{{\cal N}_v}\sum_{P_v}\delta(\varepsilon_{c\bm
  k_c} -\varepsilon_{v\bm k_v}\nonumber\\
&&\pm\hbar\Omega_{(\bm
    k_c-P_v\bm k_v)\lambda}-\hbar\omega)N_{(\bm k_c-P_v\bm
    k_v)\lambda\pm}\nonumber\\
&&\times\tilde{\cal A}^{ab}_{I;c\bm k_cv(P_v\bm k_v)\lambda}\ ,
\label{eq:X2}
\end{eqnarray}
where
\begin{equation}\tilde{\cal A}^{ab}_{I;c\bm k_cv\bm
  k_v\lambda}=\frac{1}{{\cal N}_{c,I}}\sum_{P_{c,I}}{\cal A}^{ab}_{c(P_{c,I}\bm
  k_c)v(P_{c,I}\bm k_v);\lambda}\ .
\label{eq:calA}
\end{equation}
The prime indicates the summation is only over the
irreducible wedge of the Brillouin zone, $P_{c,I}$ are the ${\cal N}_{c,I}$ symmetry
operations in $C_{4v}$ that keep the $I^{th}$ valley unchanged, while
$P_v$ are the ${\cal N}_v$ symmetry operations in $O_h$. For each process
$\{I;cv\lambda\pm\}$, the symmetry properties of the tensor with
components ${\cal
  A}^{ab}_{I;cv\lambda\pm}$ are the same as those of the tensor with
components $\tilde{\cal A}^{ab}_{I;c\bm k_cv\bm
  k_v\lambda}$. Therefore, in the $I=Z$ valley, the
$\xi_{Z;cv\lambda\pm}^{ab}$ form a second rank tensor with  only
two nonzero independent components,
\begin{eqnarray}
\xi^{xx}_{Z;cv\lambda\pm} &=& \xi^{yy}_{Z;cv\lambda\pm} \equiv \xi^{(1)}_{cv\lambda\pm}\ ,\nonumber\\
\xi^{zz}_{Z;cv\lambda\pm} &\equiv& \xi^{(2)}_{cv\lambda\pm}\ .
\label{eq:nonzeroxi}
\end{eqnarray}
Similarly, the $\zeta_{Z;cv\lambda}^{abc}$ form a third rank
pseudotensor with only two nonzero independent components,
\begin{eqnarray}
\zeta^{zxy}_{Z;cv\lambda\pm}&=& -\zeta^{zyx}_{Z;cv\lambda\pm} \equiv i\zeta^{(1)}_{cv\lambda\pm}\ ,\nonumber\\
\zeta^{xyz}_{Z;cv\lambda\pm}&=&-\zeta^{yxz}_{Z;cv\lambda\pm} \equiv i\zeta^{(2)}_{cv\lambda\pm}\ ,\nonumber\\
\zeta^{xzy}_{Z;cv\lambda\pm}&=&-\zeta^{yzx}_{Z;cv\lambda\pm} =
-i\zeta^{(2)}_{cv\lambda\pm}\ ,
\label{eq:nonzerozeta}
\end{eqnarray}
where $\xi^{(1)}$, $\xi^{(2)}$ and $\zeta^{(1)}$ are real numbers,
and $\zeta^{(2)}$ is also a real number because of inversion and time
reversal symmetry in bulk silicon. The injection coefficients in other 
valleys can be obtained by properly rotating the $Z$ valley to the corresponding
valley. 
The total injection coefficients
\begin{equation}
{\cal A}^{ab}_{cv\lambda\pm}=\sum_{I}{\cal A}^{ab}_{I;cv\lambda\pm}
\label{eq:sumvalley}
\end{equation}
have higher
symmetry, and the nonzero components satisfy
\begin{eqnarray}
   \xi^{xx}_{cv\lambda\pm} = \xi^{yy}_{cv\lambda\pm} = \xi^{zz}_{cv\lambda\pm} &=& 2\xi^{(2)}_{cv\lambda\pm} + 4\xi^{(1)}_{cv\lambda\pm}\ ,\nonumber\\
    \zeta^{xyz}_{cv\lambda\pm} = \zeta^{yzx}_{cv\lambda\pm} = \zeta^{zxy}_{cv\lambda\pm} &=&-\zeta^{xzy}_{cv\lambda\pm} =
  -\zeta^{zyx}_{cv\lambda\pm} = -\zeta^{yxz}_{cv\lambda\pm}
  \nonumber\\
  &=& i\left(2\zeta^{(1)}_{cv\lambda\pm} + 4\zeta^{(2)}_{cv\lambda\pm}\right)\ .\nonumber
\end{eqnarray}
All quantities keep the identified symmetry properties on summation of one
or several subscripts in $\{cv\lambda\pm\}$.

\section{Transitions at the band edge}
\label{sec:selectionrules}
The values of the matrix elements $\bm T_{\bar{c}\bm k_c\bar{v}\bm
  k_v,\lambda}$ at the band edge, $\bm T_{\bar{c}\bm k_c^0\bar{v}\bm
  k_v^0,\lambda}$, provide insight into the
importance of each injection process. For the indirect gap injection in
silicon, the conduction band edge in the $Z$ valley is at $\bm
k_c^0=(0,0,k_\Delta)$, while the valence band edge is at the $\Gamma$
point, $\bm k_v^0=0$.  
In analyzing transitions near the band edge, we can use the
following two approximations: i) the values of the transition matrix elements 
[Eq. (\ref{eq:transitionmatrix})] are taken at their band edge values,
and ii) the values of the wave vector, the energy, and the
phonon number of the phonons involved are all taken at their band edge values. Under 
these approximations, Eq. (\ref{eq:X2}) becomes
\begin{equation}
  {\cal A}^{ab}_{I;cv\tau\pm}(T, \omega) \approx
  \frac{2\pi}{\hbar}J_{cv}(\hbar\omega\mp\hbar\Omega_{\bm
    k_c^0\lambda})N_{\bm k_c^0\tau\pm}\bar{\cal
    A}^{ab}_{I;cv\tau}\  , 
\label{eq:selectionrule}
\end{equation}
with 
\begin{eqnarray}
J_{cv}(\hbar\omega) &=& \sum_{\bm k_c\in I,\bm
  k_v}\delta(\varepsilon_{c\bm 
  k_c}-\varepsilon_{v\bm
  k_v}-\hbar\omega)\ ,
\label{eq:jointdos}\\
\bar{\cal
  A}^{ab}_{I;cv\tau}&=&\frac{1}{{\cal N}_v}\sum_{\lambda\in\tau}\sum_{P_v}\tilde{\cal A}^{ab}_{c\bm
k_c^0v(P_v\bm k_v^0)\lambda}\ .
\label{eq:barcala}
\end{eqnarray}
Here $\tau$ indicates the phonon branches (TA, TO, LA, LO), $J_{cv}(\hbar\omega)$ is the joint density of states (JDOS) for indirect gap injection, $\bar{\cal
  A}^{ab}_{I;cv\tau}$ gives the symmetrized transition matrix
elements at the band edge, and $\sum_{\lambda\in\tau}$ indicates summation over all modes in the $\tau^{th}$ branch. Under
the parabolic band approximation\cite{Phys.Rev._127_765_1962_Hartman}, the JDOS is
\begin{equation}
J_{cv}(\hbar\omega)\propto
(E_{ig}-\hbar\omega)^2\ .
\label{eq:lineshape}
\end{equation}

We now turn to $\bar{\cal
  A}^{ab}_{I;cv\tau}$. Because
the HH and LH bands are degenerate at the $\Gamma$ point and their wave
functions are not uniquely determined, unambiguous values of ${\cal A}^{ab}_{c\bm k_c^0v\bm k_v^0\lambda}$ and $\tilde{\cal
  A}^{ab}_{c\bm k_c^0v\bm k_v^0\lambda}$ for the HH and LH bands
separately do not exist at the band edge. But 
this is not a problem for $\bar{\cal A}^{ab}_{I;cv\tau}$, due to the
summation over all symmetry operations. To show this clearly, we indicate
all 
intermediate states in the transition matrix elements implicitly by
writing
  \begin{equation}
    \bm T_{\bar{c}\bm k_c\bar{v}\bm k_v;\lambda} = \frac{e}{\omega}\langle \bar{c}
    \bm k_c|\hat{\bm T}_{c\bm k_cv\bm
      k_v;\lambda}|\bar{v}\bm k_v\rangle\ ,
    \label{eq:selectiont}
    \end{equation}
with the operator 
\begin{eqnarray}
\hat{\bm T}_{c\bm k_cv\bm k_v;\lambda}  &\equiv&
H^{ep}_{\lambda}(\bm k_c-\bm k_v)\frac{1}{\hbar\omega-H_e+\varepsilon_{v\bm 
    k_v}}\hat{\bm v}  \nonumber\\
&+& \hat{\bm v} \frac{1}{\varepsilon_{c\bm 
    k_c}-H_e-\hbar\omega}H^{ep}_{\lambda}(\bm k_c-\bm k_v)
\end{eqnarray}
Substituting Eq. (\ref{eq:selectiont}) into
Eq. (\ref{eq:transitionmatrix}), it is easy to find that the
expression for $\bar{\cal
  A}^{ab}_{I;cv\tau}$ includes a summation ${{\cal N}_v}^{-1}\sum_{P_v}P_v|\bar{v}\bm k_v^0\rangle\langle
\bar{v}\bm k_v^0|P_v$, which equals $\sum_{\bar{v}^{\prime}\in
  \mathtt{LH,HH}}|\bar{v}^{\prime}\bm k_v^0\rangle\langle \bar{v}^{\prime}\bm
k_v^0|/4$ when $v$ is the HH or LH band. So $\bar{\cal
  A}^{ab}_{I;c\mathtt{HH}\tau}=\bar{\cal
  A}^{ab}_{I;c\mathtt{LH}\tau}=\left(\bar{\cal
  A}^{ab}_{I;c\text{HH}\tau}+\bar{\cal
  A}^{ab}_{I;c\mathtt{LH}\tau}\right)/2$ are unambiguous, and give the same
value for both the HH and LH bands. This conclusion mirrors a similar
one in the study of injection across the direct gap\cite{Phys.Rev.B_76_205113_2007_Nastos}. In the following, we
focus on selection rules for $\bar{\cal A}^{ab}_{Z;cv\lambda}$. Li and
Dery \cite{Phys.Rev.Lett._105_037204_2010_Li} also discussed
the selection rules in the context of luminescence by considering only
the lowest conduction band and the highest valence band as the
intermediate states; they also assume equal amplitudes for the two
interference processes shown in Fig.~\ref{fig:bandstructure}. Here we
give a general discussion using Eq. (\ref{eq:selectiont}), without relying on the properties of the
intermediate states.

Without spin-orbit coupling, the valence band 
states at the $\Gamma$ point transform according to the representation
$\Gamma_{25}^{\prime}$ (with basis functions that transform as $\{yz, zx, xy\}$, which are labelled as 
$\{{\cal X}, {\cal Y}, {\cal Z}\}$
here)\cite{Phys.Rev.B_76_205113_2007_Nastos}. We first consider the electron states that lie in the $Z$
valley, in which the conduction band edge state transforms according
to the representation $\Delta_1$ (basis function $\{z\}$)\cite{peteryu}.
The phonon states involved transform according to the representation
$\Delta_1$ (basis function $\{z\}$) for the LA phonon mode,
$\Delta_2^{\prime}$ (basis function $\{x^2-y^2\}$) for the LO phonon mode and
$\Delta_5$ (basis functions $\{x,y\}$) for the TA/TO phonon modes;\cite{peteryu}
$H^{ep}_{\lambda}(\bm k_c^0)$ has the same symmetry properties 
as the $\lambda^{th}$ branch phonon polarization vector. The velocity $\hat{\bm
  v}$ transforms according to the representation
$\Delta_1\oplus\Delta_2$. All nonzero components of the vector $\bm T_{c\bm
  k_c^0 v\bm  k_v^0\lambda}$ are listed in Table\ \ref{tab:nz-one} for
the different valence bands and phonon modes.  In Li and Dery's approximations,
$T_2=2T_1$ and $T_2^{\prime}=2T_1^{\prime}$ because the two transition processes summed are of
equal magnitude,  and $T_5$ vanishes due to their limitation of the
intermediate states.  In contrast, we keep all nonzero terms in our
treatment, and give their values in Sec.~\ref{sec:epm}.
\begin{table}[htp]
  \centering
  \begin{tabular}[t]{|c|c|c|c|c|}
    \hline
    \hline
    \multicolumn{1}{|c|}{}&\multicolumn{4}{c|}{phonon mode $\lambda$}\\
\cline{2-5}
    \multicolumn{1}{|c|}{${\bm
        T}_{c\bm k_c^0v\bm k_v^0\lambda}$} &
      \multicolumn{2}{|c|}{TA/TO
       } & LA & LO \\
    \cline{2-5}
    \multicolumn{1}{|c|}{}&$x$&$y$&$z$&$x^2-y^2$\\
    \hline
    $|{\cal X}\rangle$ & 0 &$(0,0,T_1^{(\prime)})$  & $(0,T_3,0)$ & $(T_4,0,0)$\\
    \cline{1-5}
    $|{\cal Y}\rangle$ & $(0,0,T_1^{(\prime)})$ & 0 & $(T_3,0,0)$&$(0,T_4,0)$\\
    \cline{1-5}  
    $|{\cal Z}\rangle$ & 
    $(0,T_2^{(\prime)},0)$ & $(T_2^{(\prime)},0,0)$ & 0& $(0,0,T_5)$ \\
    \cline{1-5}
    \hline
  \end{tabular}
  \caption{The transition matrix elements ($T^x_{c\bm k_c^0,v\bm
      k_v^0,\lambda}$, $T^y_{c\bm k_c^0,v\bm k_v^0,\lambda}$,$T^z_{c\bm
      k_c^0,v\bm k_v^0,\lambda}$). The symbols with superscript
    prime are for TO phonons. All quantities can be taken as real numbers.}
  \label{tab:nz-one}
\end{table}

With spin-orbit coupling, the valence bands at the $\Gamma$ point are split into HH
($|\frac{3}{2},\pm\frac{3}{2}\rangle$), LH
($|\frac{3}{2},\pm\frac{1}{2}\rangle$) and SO
($|\frac{1}{2},\pm\frac{1}{2}\rangle$) bands with the following
states:
{\allowdisplaybreaks
\begin{eqnarray}
  |\frac{1}{2},+\frac{1}{2}\rangle &=
  &\frac{1}{\sqrt{3}}|{\cal Z}\rangle|\uparrow\rangle +
  \frac{1}{\sqrt{3}}|{\cal X} +
  i{\cal Y}\rangle|\downarrow\rangle\ ,\nonumber\\ 
  |\frac{1}{2},-\frac{1}{2}\rangle &=
  &\frac{1}{\sqrt{3}}|{\cal X} -
  i{\cal Y}\rangle|\uparrow\rangle-\frac{1}{\sqrt{3}}|{\cal Z}\rangle|\downarrow\rangle\ ,\nonumber\\
  |\frac{3}{2},+\frac{1}{2}\rangle &=
  &\sqrt{\frac{2}{3}}|{\cal Z}\rangle|\uparrow\rangle -
  \frac{1}{\sqrt{6}}|{\cal X} +i{\cal Y}\rangle|\downarrow\rangle\
  ,\nonumber\\ 
  |\frac{3}{2},-\frac{1}{2}\rangle &=
  &\frac{1}{\sqrt{6}}|{\cal X}-i{\cal Y}\rangle|\uparrow\rangle +
  \sqrt{\frac{2}{3}}|{\cal Z}\rangle|\downarrow\rangle\ ,\nonumber\\ 
  |\frac{3}{2},+\frac{3}{2}\rangle &=
  &-\frac{1}{\sqrt{2}}|{\cal X}+i{\cal Y}\rangle|\uparrow\rangle\ ,\nonumber\\
  |\frac{3}{2},-\frac{3}{2}\rangle &=
  &\frac{1}{\sqrt{2}}|{\cal X} - i{\cal Y}\rangle|\downarrow\rangle\ .
\end{eqnarray}
}

The SO band at the $\Gamma$
point is lower than the degenerate HH and
LH bands by $44$\ meV. At the conduction band edge, the states can be
approximately written as $|z\uparrow\rangle$ and
$|z\downarrow\rangle$ due to the very small spin
mixing. The transition matrix elements between these states with
spin-orbit coupling can be obtained by linearly combining the terms in
Table~\ref{tab:nz-one}. 
\begin{figure}[htp]
  \centering
  \includegraphics[height=7cm]{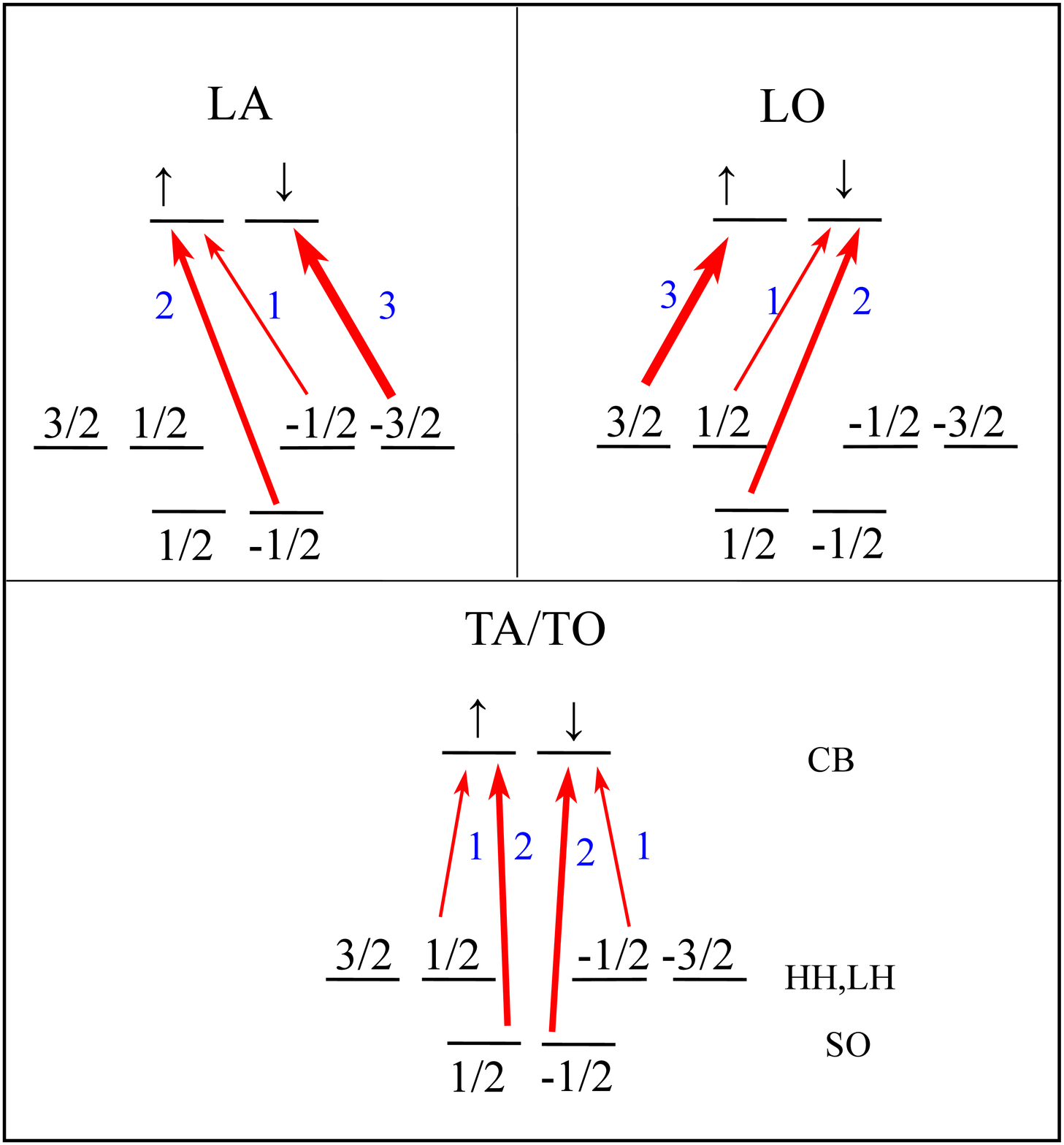}
  \caption{(color online). Diagram for optical indirect
    transitions from the band edge of the valence bands (HH, LO, and
    SO) to the band edge of the conduction band (CB) in the $Z$
    valley, under $\sigma^-$ light. Both the hole and electron states
    are quantized along the $z$ direction. The red arrows stand for allowed
    transitions, whose probabilities are proportional to the product
    of the numbers next to the arrows and factors of
    $\frac{1}{3}T_3^2$, $\frac{1}{3}T_4^2$, and
    $\frac{1}{3}[T_2^{(\prime)}]^2$ for LA, LO, and TA/TO
    phonon-assisted processes, respectively.}
  \label{fig:selectionrule}
\end{figure}

For $\sigma^-$ light, $\bm E=(\hat{\bm x}-i\hat{\bm y})/\sqrt{2}$, we show in
Fig.~\ref{fig:selectionrule} the possible optical transitions from the
valence bands to the $Z$ valley of the conduction band. Here the spin quantization
directions of both electrons and holes are chosen along the $z$
direction. It is obvious that the phonon states play a key role in 
these transitions. For transitions from HH and LH bands, the LA
and LO phonon-assisted processes inject spin polarization along the $-z$
and $z$ directions, respectively, with a DSP of $50\%$, yet there is no spin
polarization from the TA/TO phonon-assisted processes. Despite its
spin independence, the electron-phonon interaction still affects the selection rules for spin
injection since all states involved are not pure spin states due to
spin-orbit coupling. Therefore, it is not adequate to treat the
indirect transitions as a spin dependent virtual optical transition
combined with a phonon emission or absorption process that does not affect the
spin\cite{Proc.9thInt.Coef.Phys.Semicond._2_1139_1968_Lampel,PhdThesis_Verhulst_2004}. 

\begin{figure}[htp]
  \centering
  \includegraphics[height=7cm]{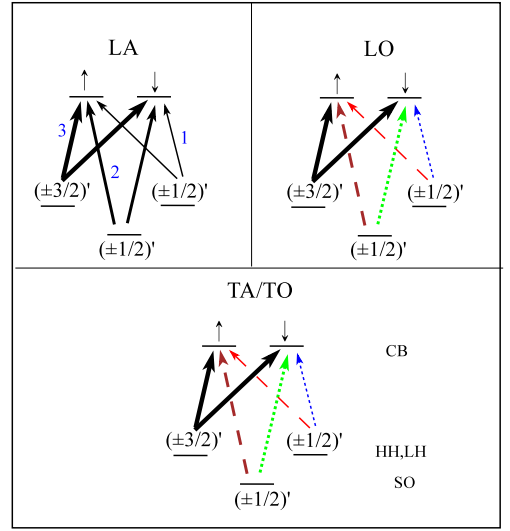}
  \caption{(color online). Diagram for optical indirect
    transitions from the band edge of the valence bands (HH, LO, and
    SO) to the band edge of the conduction band (CB) in the $X$
    valley, under $\sigma^-$ light. The electron states are quantized
    along the $z$ direction, and the prime beside the hole states indicate that they are quantized
    along the $x$ direction. The transition strength for each
      color arrows in each phonon branch is as follows: In LA-assisted transitions, the factor is
    $\frac{1}{24}T_3^2$ for black arrows. In LO-assisted transitions, the factors are
    $\frac{1}{8}T_4^2$ for black arrows,
    $\frac{1}{24}(T_4\pm2T_5)^2$ for blue thin dotted ($+$) and red
    thin dashed ($-$) arrows,
    and $\frac{1}{12}(T_4\pm T_5)^2$ for brown thick dashed ($+$) and
    green thin dotted ($-$)
    arrows. In TA/TO-assisted transitions, the factors are
    $\frac{1}{4}[T_1^{(\prime)}]^2$ for black arrows,
    $\frac{1}{12}[(T_1^{(\prime)})^2 + 2(T_2^{(\prime)})^2 \pm 
    2T_2^{(\prime)}T_1^{(\prime)}]$ for red thin dashed ($+$) and
    blue thin dotted  ($-$) arrows, and $\frac{1}{12}[2(T_1^{(\prime)})^2 + (T_2^{(\prime)})^2 \pm 
    2T_2^{(\prime)}T_1^{(\prime)}]$ for green thick dotted
    ($+$) and brown thick dashed ($-$) arrows.
  } 
  \label{fig:selectionrule2}
\end{figure}
Now we look at the transitions to the electron states in the $X$
valley. The possible optical transitions are complicated, and are shown in
Fig.~\ref{fig:selectionrule2}. To simplify the diagram, we choose the
quantization axis of the hole states along the $x$ direction, and that
of the electron states along the $z$ direction. Here the LO and TA/TO
phonons can play a role in spin injection, but the LA phonon
cannot. Thus there is strong valley anisotropy in the injection of spins. 
In Li and
Dery's approximations, the TA/TO phonon-assisted processes give a DSP of $1/3$ 
from the HH and LH bands, and the LO phonon-assisted process gives no spin polarization.
In our more detailed analysis, we cannot simply identify a DSP for
each process, as some of these
processes are determined by more than one nonzero parameter. We
discuss the actual values of these at the end of section \ref{sec:epm}.

Because the band edge transitions strongly depend on the choice of the
electron and hole states, it is constructive to give the nonzero components of
$\bar{\cal A}^{ab}_{Z;cv\tau}$. They can be  specified by giving
the values of $\bar{\cal A}^{(1,2)}_{cv\tau}$, in terms of which
$\bar{\cal A}^{ab}_{Z;cv\tau}$ are found following the pattern of
Eq.~(\ref{eq:nonzeroxi}) and (\ref{eq:nonzerozeta}).  The results are shown in
Table~\ref{tab:nz-injection}. In the calculation, we have
used the result that  the contributions for the HH and LH bands at the
$\Gamma$ point are exactly the same. 
\begin{center}
  \begin{table}[htp]
    \begin{tabular}[t]{|c|c|c|c|c|}
      \hline
      \multicolumn{2}{|c|}{$\tau$}&\multicolumn{1}{c|}{TA/TO} & \multicolumn{1}{c|}{LA
        } & \multicolumn{1}{c|}{LO
        }\\
      \hline
      & SO &  $(0,-\frac{2}{3}T_1^{(\prime)}T_2^{(\prime)})$ 
      & $(-\frac{2}{3}T_3^2,0)$ 
      & $(\frac{2}{3}T_4^2,\frac{2}{3}T_4T_5)$\\
      \cline{2-5}
      \raisebox{2.0ex}[0pt]{$\bar{\zeta}$}& LH,HH
      &$(0,\frac{1}{3}T_1^{(\prime)}T_2^{(\prime)})$
      &$(\frac{1}{3}T_3^2,0)$ 
      &$(-\frac{1}{3}T_4^2,-\frac{1}{3}T_4T_5)$
      \\
      \hline
      \hline
      &SO
      & $(\frac{2}{3}[T_2^{(\prime)}]^2,\frac{4}{3}[T_1^{(\prime)}]^2)$
      &$(\frac{2}{3}T_3^2,0)$ 
      &$(\frac{2}{3}T_4^2,\frac{2}{3}T_5^2)$
      \\
      \cline{2-5}
      \raisebox{2.0ex}[0pt]{$\bar{\xi}$}&LH,HH
      &$(\frac{2}{3}[T_2^{(\prime)}]^2,\frac{4}{3}[T_1^{(\prime)}]^2)$
      &$(\frac{2}{3}T_3^2,0)$
      &$(\frac{2}{3}T_4^2,\frac{2}{3}T_5^2)$ \\
      \hline
    \end{tabular}
    \caption{Band edge transition rates $(\bar{\cal
        A}^{(1)}_{cv\tau},\bar{\cal A}^{(2)}_{cv\tau})$ with $\cal A$
      being $\xi$ for carrier injection and $\zeta$ for spin injection.}
    \label{tab:nz-injection}
  \end{table}
  \end{center}    

From Table\ \ref{tab:nz-injection}, we find that some results are
similar to those for direct gap injection: The band edge transition
magnitudes for carrier injection are the same for each valence band,
while the ones for spin injection are only identical for HH and LH
bands, and satisfy
\begin{equation}
  \sum_{v}\bar{\zeta}^{(1,2)}_{cv\tau} = 0\ . 
\label{eq:summation}
\end{equation}
The vanishing sum does not mean that the spin polarization becomes
zero when the laser pulse is wide enough to involve all valence bands, 
because the densities of states for different valence bands are
different. However, as the photon energy increases, the optical
transition occurs away from the band edge, where the involved
electron and hole states are superpositions of the band edge states of
all valence bands. As a consequence of Eq.~(\ref{eq:summation}),
important band mixing leads to a small DSP.

\section{Model for Electron states and phonon states}
\label{sec:epm}
To look at the carrier and spin injection away from the band edge, a
full band structure model of the electron and phonon states is
necessary. We use the
EPM\cite{Phys.Rev.B_10_5095_1974_Chelikowsky,Phys.Rev.B_14_556_1976_Chelikowsky,Phys.Rev._149_504_1966_Weisz} 
for electron states and the ABCM \cite{Phys.Rev.B_15_4789_1977_Weber} for phonon
states. We describe these now, and give the resulting electron-phonon interaction for indirect
gap injection. In the EPM, electrons are described by the pseudo-Hamiltonian, 
\begin{equation}
  H_e = \frac{{\bm p}^2}{2m_0} + \sum_{i\alpha}v(\bm r-\bm
  R_{i\alpha})\ .
\label{eq:hamiltonian}
\end{equation}
Here $\bm p$ is the momentum operator; $\bm R_{i\alpha}=\bm R_{i} +
\bm \tau_{\alpha}$ is the equilibrium position for the $\alpha^{th}$
($\alpha=1,2$) atom in the $i^{th}$ primitive cell located at $\bm R_i$
with $\bm{\tau}_{1}=0$ and $\bm{\tau}_2=\frac{a}{4}(1,1,1)$, and $a$ is
the lattice constant; $v(\bm r, \bm p) = v_{\mathtt{L}}(\bm
r) + v_{\mathtt{NL}}(\bm
r, \bm p) + v_{\mathtt{so}}(\bm r, \bm p)$ is the atomic empirical pseudopotential,
in which $v_{\mathtt{L}}(\bm r)$ and $v_{\mathtt{NL}}(\bm r, \bm p)$ are the
local and nonlocal spin independent pseudopotentials, 
respectively\cite{Phys.Rev.B_10_5095_1974_Chelikowsky,Phys.Rev.B_14_556_1976_Chelikowsky},
and $v_{\mathtt{so}}(\bm r, \bm p)$ is a spin dependent contribution \cite{Phys.Rev._149_504_1966_Weisz} 
introduced to fit the spin split-off energy
at the $\Gamma$ point\cite{Phys.Rev.Lett._104_016601_2010_Cheng}. Eigenstates of $H_e$ are
given by the Bloch states $|n\bm k\rangle =\sum_{\bm g}c_{n\bm k}(\bm
g)|\bm k+\bm g\rangle$, where $|\bm k\rangle$ are plane wave states,
$\bm g$ is a reciprocal lattice vector, and $c_{n\bm  k}=\begin{pmatrix}c_{n\bm k}^{\uparrow}&c_{n\bm
    k}^{\downarrow}\end{pmatrix}^T$ are coefficients obtained by solving
the single-particle Schr\"odinger equation. 

The Fourier
transform of the pseudopotential $v(\bm r, \bm p)$  is 
\begin{equation}
v(\bm k_1, \bm k_2)=\int \frac{d\bm r}{a^3}
e^{-i\bm k_1\cdot\bm r}v(\bm r, \bm p)e^{i\bm k_2\cdot\bm r}\ .
\end{equation}
For the local pseudopotential $v_{\mathtt{L}}(\bm r)$ taken to be of
the form $v_{\mathtt{L}}(r)$ with $r=|\bm r|$, we have that $v_{\mathtt{L}}(\bm k_1,\bm
k_2)$ depends only on $|\bm k_1-\bm k_2|$, and we write it as
$v_{\mathtt{L}}(|\bm k_1-\bm k_2|)$. Chelikowsky and Cohen
\cite{Phys.Rev.B_10_5095_1974_Chelikowsky,Phys.Rev.B_14_556_1976_Chelikowsky}
showed that a suitable electron band structure can be produced by
taking the values of $v_{\mathtt{L}}(k)$ only at $k^2=3$, 4, and 11
$(2\pi/a)^2$, which lead to the conduction band energy
at the $X$ point of 1.17~eV. In order to produce the direct band gap
$E_g$, the indirect band gap $E_{ig}$, and the location of the
conduction band edge $\bm k_c^0$ correctly after including the
spin-orbit coupling, we use the following parameters 
$v_{\mathtt{L}}(k) = -3.496$, $-0.544$, $0.437$, $0.429$, and
$0.1373$~eV for $k^2=3$, 4, 11, 16, and 19 $(2\pi/a)^2$. 
The  calculated band structure is shown in
Fig.~\ref{fig:bandstructure}. The calculation is for zero
temperature. With increasing temperature, the electron-phonon interaction induces changes at the band edge and
shifts the indirect and direct band  gaps\cite{Phys.Rev.B_31_2163_1985_Lautenschlager,J.Appl.Phys._51_2634_1980_Mahan,J.Appl.Phys._45_1846_1974_Bludau,J.Appl.Phys._79_6943_1996_Alex}.
For silicon, the shift in the conduction band edge is minor; we
absorb it into $E_{ig}$ in  the following. 

For the phonons, the ABCM is used for calculating the polarization
vectors $\bm{\epsilon}_{\bm  q\lambda}^\alpha$ and energies $\hbar\Omega_{\bm q\lambda}$. The
calculated band structure fits the experiments
well\cite{Phys.Rev.B_15_4789_1977_Weber}. The energies of phonons with
wave vector $\bm k_c^0$, which are involved in the band edge transition, are $\hbar\Omega_{\bm
  k_c^0,\mathtt{TA}}=19$~meV, $\hbar\Omega_{\bm
  k_c^0,\mathtt{LA}}=43$~meV, $\hbar\Omega_{\bm
  k_c^0,\mathtt{LO}}=53$~meV , and $\hbar\Omega_{\bm k_c^0,\mathtt{TO}}=57$~meV.

By shifting the atom position from the equilibrium position $\bm
R_{i\alpha}$ to $\bm R_{i\alpha}+\bm u_{i\alpha}$, and then expanding
the electron Hamiltonian (\ref{eq:hamiltonian}) to linear
order in $\bm u_{i\alpha}$, we identify the electron-phonon interaction
 as $H^{ep} =
-\sum_{i\alpha}\bm{\nabla} v(\bm r-\bm
R_{i\alpha})\cdot\bm u_{i\alpha}$. The atomic displacement is
usually expanded by the phonon polarization vectors as 
\begin{equation}
\bm u_{i\alpha}=\sum\limits_{\bm q\in
  1^{st}\mathtt{BZ},\lambda}\left(\frac{\hbar}{\rho\Omega_{\bm
      q\lambda}}\right)^{1/2}(a_{\bm q\lambda}+a_{-\bm
  q\lambda}^{\dagger})\bm{\epsilon}_{\bm q\lambda}^\alpha e^{i\bm
  q\cdot\bm R_i}\ ,
\end{equation}
with $\rho$ being the mass density of silicon. The transition matrix elements between
different electron states are given as
\begin{eqnarray}
  &&M_{n_1\bm k_1n_2\bm k_2,\lambda}=  -i\sqrt{\frac{\hbar}{\rho\Omega_{\bm q\lambda}}}\sum_{\bm
  g_1\bm g_2}\bigg[\Delta\bm
k\cdot\sum_{\alpha}\bm{\epsilon}_{\bm q,\alpha}^{\lambda}e^{-i\Delta\bm 
      k\cdot\bm{\tau}_{\alpha}}\bigg]\nonumber\\
&&\hspace{1cm}\times c^{\dag}_{n_2\bm k_2}(\bm
g_1)v(\bm  k_1+\bm g_1,\bm k_2+\bm g_2)c_{n_2\bm k_2}(\bm
g_2)\ ,
\label{eq:ep}
\end{eqnarray}
where $\bm q=\bm k_1-\bm k_2$,  $\Delta\bm k=\bm
k_1+\bm g_1-\bm k_2-\bm g_2$.

In calculating the matrix elements of the electron-phonon interaction
given in Eq.~(\ref{eq:ep}), the values of $v_{\mathtt{L}}(k)$ at all $k$ points are necessary, and
they are obtained using the natural cubic spline interpolation
\cite{Phys.Rev.B_48_14276_1993_Rieger} on the discrete points given
above and two further restrictions: the first is the value
$v_{\mathtt{L}}(0)$; the second is a cut off at high $k$,
$v_{\mathtt{L}}(k>k_{cut})=0$, where $k_{cut}$ is a cut-off value. Bednarek and R\"ossler
\cite{Phys.Rev.Lett._48_1296_1982_Bednarek} showed that the values
assumed for $v_{\mathtt{L}}(0)$ and $k_{cut}$ strongly affect the
calculated matrix elements of the electron-phonon interaction in Eq.~(\ref{eq:ep}). In our 
calculation, we find that the assumed value of $k_{cut}$ doesn't
significantly affect $H^{ep}$ for
$k_{cut}>3k_F$, and we set $k_{cut}=3k_F$. We show
the dependence of indirect absorption on $v_{\mathtt{L}}(0)$ in Table
\ref{tab:bevalue1} and \ref{tab:bevalue2} by taking
$v_{\mathtt{L}}(0) = -\frac{2}{3}E_F$ (denoted as case A) and 
$v_{\mathtt{L}}(0) = 0$ (case B). Here $k_F$ and $E_F$ are the Fermi wave
vector and Fermi energy of the free electron gas appropriate to the
valence electron density in silicon. 

\begin{widetext}
\begin{center}
  \begin{table}[htp]
    \begin{tabular}[t]{|c|c|c|c|c|c|c|c|c|c|c|c|c|}
      \hline
      \multicolumn{3}{|c|}{}&\multicolumn{3}{|c|}{TA}&\multicolumn{3}{c|}{TO} & \multicolumn{1}{c|}{LA} & \multicolumn{3}{c|}{LO}\\
      \cline{4-13}
      \multicolumn{3}{|c|}{}&$T_1^2$&$T_2^2$&$T_1T_2$&$(T_1^{\prime})^2$&$(T_2^{\prime})^2$&$T_1^{\prime}{T_2^{\prime}}$&$T_3^2$&$T_4^2$&$T_5^2$&$T_4T_5$\\
      \hline
      \multicolumn{3}{|c|}{A}&0.020&0.018&$-0.019$&0.131&1.000&$-0.358$&0.018&0.218&0.001&$-0.014$\\
      \hline
      \multicolumn{3}{|c|}{B}&0.066&0.090&$-0.077$&0.120&1.024&$-0.348$&0.020&0.192&0.026&$-0.070$\\
      \hline
    \end{tabular}
    \caption{Relative band edge values of all quantities listed in Table
      \ref{tab:nz-injection} for the phonon emission process. The photon
      energies for different phonon branches are taken as
      $\hbar\omega=E_{ig}+\hbar\Omega_{\bm k_c^0\lambda}$. All values
      are normalized with respect to the value of $|T_2^{\prime}|^2$
      in case A. Its value is
      $(T_2^{\prime})^2=1.8\times10^{-86}$~J$^2$V$^{-2}$m$^{-1}$.}
    \label{tab:bevalue1}
  \end{table}
  \begin{table}[htp]
  \begin{tabular}[t]{|c|c|c|c|c|c|c|c|c|c|c|}
      \hline
      \multicolumn{2}{|c|}{}
      &\multicolumn{3}{|c|}{$Z$ valley}
      &\multicolumn{3}{|c|}{$X$ valley}
      &\multicolumn{3}{|c|}{total}\\
      \cline{3-11}
      \multicolumn{2}{|c|}{\raisebox{2.0ex}[0pt]{}}&$c^Z\equiv$&$s^Z\equiv$&$\mathtt{DSP}^{Z}\equiv$&$c^X\equiv$&$s^X\equiv$&$\mathtt{DSP}^X\equiv$&$c\equiv$&$s\equiv$&$\mathtt{DSP}\equiv$\\
      \multicolumn{2}{|c|}{\raisebox{2.0ex}[0pt]{}}&$\bar{\xi}_{c\text{HH}\lambda+}^{(1)}$&$-\frac{2}{\hbar}\bar{\zeta}_{c\text{HH}\lambda+}^{(1)}$&$s^Z/c^Z$&$\frac{1}{2}(\bar{\xi}_{c\text{HH}\lambda+}^{(1)}+\bar{\xi}_{c\text{HH}\lambda+}^{(1)})$&$-\frac{2}{\hbar}\bar{\zeta}_{c\text{HH}\lambda+}^{(2)}$&$s^X/c^X$&$4c^X+2c^Z$&$4s^X+2s^Z$&$s/c$\\
      \hline
      &TA&0.0117&0&0&0.019&$-0.006$&$-32\%$&0.100&$-0.025$&$-25\%$\\
      \cline{2-11}
      &LA&0.0122&$-0.0062$&$-51\%$&0.006&0.00022&$4\%$&0.0488&$-0.0115$&$-24\%$\\
      \cline{2-11}
      \raisebox{2.0ex}[0pt]{A}&LO&0.145&0.0727&$50\%$&0.073&0.00473&$6\%$&0.583&0.164&$28\%$\\
      \cline{2-11}
      &TO&0.667&$-0.0056$&$-1\%$&0.42&$-0.119$&$-28\%$&3.01&-0.489&$-16\%$\\
      \hline
      \hline
      &TA&0.0603&$-0.00045$&$-1\%$&0.074&$-0.026$&$-35\%$&$0.42$&$-0.103$&-25\%\\
      \cline{2-11}
      &LA&0.0135&$-0.0067$&$-50\%$&$0.0068$&0&0&$0.054$&$-0.013$&$-24\%$\\
      \cline{2-11}
      \raisebox{2.0ex}[0pt]{B}&LO&0.128&0.064&50\%&0.072&0.023 &32\%&$0.546$&$0.222$&41\%\\
      \cline{2-11}
      &TO&0.682&0&0&0.421&$-0.116$&$-28\%$&3.05&$-0.475$&$-16\%$\\
      \hline
    \end{tabular}
    \caption{Band edge values of $\bar{\cal
        A}^{(1,2)}_{c\text{HH}\lambda}$. The classification in $Z$ and
      $X$ valleys follows Eq.~(\ref{eq:coefpresent}). All values 
      are normalized with respect to the value of $|T_2^{\prime}|^2$
      in case A.}
    \label{tab:bevalue2}
  \end{table}
\end{center}
\end{widetext}

For carrier injection, the contributions from the phonon branches
other than TA are essentially the same in case A and B, while for spin
injection the phonon processes involving TA or LO phonons and electron
injection into the $X$ valley give contributions that depends strongly
on the value of $v_{\mathtt{L}}(0)$. The DSP for the LO phonon-assisted
process is about $6\%$ in case A and $32\%$ in case B. Nevertheless, by far the most
important contribution to indirect gap injection comes from the TO phonon-assisted
process, which is almost independent of $v_{\mathtt{L}}(0)$ for both
carrier and spin injection. In the following, we use the parameters
of case A.

Contrary to what was assumed in Li and Dery's results, our results
show that $T_2=2T_1$ or $T_2^{\prime}=2T_1^{\prime}$ cannot be
well satisfied. But the calculated DSP for TA/TO phonon-assisted
processes in the $X$ valley is still close to $1/3$. The value of $T_5$ is
reasonably small.

\section{Calculations}
\label{sec:cal}
To obtain the injection rates, we can focus only on energies near the
band edge and rely on Eq.~(\ref{eq:selectionrule}), or numerically
evaluate Eq.~(\ref{eq:X2}) using the results of a full band structure
calculation. Both strategies require a six-dimensional
integration over the electron and hole wavevectors ranging within the
BZ. In the present work, we vary the photon energy $\hbar\omega$ to
about 1.5~eV above $E_{ig}$, which results in an effective integration
volume consisting of about 3/8 of the volume of the whole BZ, and a
very demanding calculation. Similar to the integration used in the
calculation of direct gap injection, the integrand here is composed of an energy 
conservation term, {\it i.e.} the Dirac $\delta$ function, and a
transition matrix element term. In a calculation of direct gap
injection, the $\delta$ function is evaluated using a
LATM\cite{Phys.Rev.B_49_16223_1994_Blochl,Phys.Rev.B_76_205113_2007_Nastos}
on a very fine grid, and the transition matrix elements are
calculated on each grid point. However, this method is not practical
in a calculation of indirect gap injection, because the transition matrix elements
 are too complicated to be evaluated on each
point of a fine grid. Instead, we find that we can obtain a
converged result by using separate grids for these two terms, adopting
a finer grid for the Dirac $\delta$ function and interpolating the
transition matrix elements on a rougher grid. The details of this
method are in Appendix \ref{app:numerical}.

In our calculation, the valence bands include HH, LH, and
SO bands, the conduction bands include the lowest and
the first excited conduction bands, and the intermediate states are
chosen as the lowest 30 bands to ensure convergence. 

The total carrier and spin injection from any pair of valence
bands via an emission or absorption process involving a phonon of any
branch, as well as the carrier and spin
injection into any conduction band valley, are completely determined once the quantities
$\xi^{(1,2)}_{cv\lambda\pm}$ and $\zeta^{(1,2)}_{cv\lambda\pm}$ are
specified: The nonvanishing Cartesian tensor components follow from
Eqs.~(\ref{eq:nonzeroxi}) and (\ref{eq:nonzerozeta}), the sum over the
different valleys follows from Eq.~(\ref{eq:sumvalley}), and the full response
tensors follow from Eq.~(\ref{eq:totala}); once these are determined the
injection rates can be calculated from Eq.~(\ref{eq:injection}) for
any light polarization. Instead of presenting our calculated
results for $\xi^{(1,2)}_{cv\lambda\pm}$ and
$\zeta^{(1,2)}_{cv\lambda\pm}$ below, we present instead 
{\allowdisplaybreaks
\begin{eqnarray}
\xi^{(X)}_{cv\lambda\pm}&\equiv&\frac{1}{2}[\xi_{cv\lambda\pm}^{(1)}+\xi_{cv\lambda\pm}^{(2)}]\
,\nonumber\\
\xi^{(Z)}_{cv\lambda\pm} &\equiv& \xi_{cv\lambda\pm}^{(1)}\ ,\nonumber\\
\zeta^{(X)}_{cv\lambda\pm}&\equiv&-\zeta_{cv\lambda\pm}^{(2)}\
,\nonumber\\
\zeta^{(Z)}_{cv\lambda\pm}&\equiv&-\zeta_{cv\lambda\pm}^{(1)}\ .
\label{eq:coefpresent}
\end{eqnarray}}
These are the terms that appear in a simple excitation scenario using
$\sigma^-$ light, as
described in the following section. They correspond to the injection into
different valleys via the different processes. Nonetheless, we stress
that given the quantities in Eq.~(\ref{eq:coefpresent}), we can
construct the carrier and spin injection in Eq.~(\ref{eq:injection}) for any
polarization using Eqs.~(\ref{eq:nonzeroxi}), (\ref{eq:nonzerozeta}), and (\ref{eq:totala}).

\section{Results}
\label{sec:result}
In the following, we focus on injection under $\sigma^-$
light. In this case, the six
valleys can be divided into two sets: \{$Z,\bar{Z}$\} and
\{$X,\bar{X},Y,\bar{Y}$\}. The injection is identical for all the
valleys within each set. The carrier and spin injection from the
valence band $v$ into the conduction band $c$ in the $I^{th}$ valley
via a $\lambda^{th}$-branch phonon emission ($+$) or absorption ($-$)
are identified by $\dot{n}_{I;cv\lambda\pm}$ and
$\dot{\bm S}_{I;cv\lambda\pm}$, respectively,
which are given from Eq.~(\ref{eq:coefpresent}) as
\begin{eqnarray}
  \dot{n}_{I;cv\lambda\pm}&=&\xi^{(I)}_{cv\lambda\pm} |E_0|^2\ ,\nonumber\\
  \dot{S}^z_{I;cv\lambda\pm}&=&\zeta^{(I)}_{cv\lambda\pm}|E_0|^2\ ,
\end{eqnarray}
with $\dot{S}^{x/y}_{I;cv\lambda\pm} =0$. Accordingly, the
coefficients 
\begin{eqnarray*}
  \xi&=&\sum_{I;cv\lambda\pm}\xi^{(I)}_{cv\lambda\pm},\quad \quad\ \xi^{(I)}=\sum_{cv\lambda\pm}\xi^{(I)}_{cv\lambda\pm}, \\
  \xi^{(I)}_{\tau}&=&\sum_{cv\pm;\lambda\in\tau}\xi^{(I)}_{I;cv\lambda\pm},\quad  \xi^{(I)}_{cv\tau}=\sum_{\pm;\lambda\in\tau}\xi^{(I)}_{cv\lambda\pm}
\end{eqnarray*}
are also used in the following to understand the injection properties. Here
$\tau=$ TA, LA, LO, TO are the phonon branches; as before, $\sum_{\lambda\in\tau}$
indicates to a sum over all phonon modes in the $\tau^{th}$ branch. Similar
notation is used for the injection coefficients of spins. 
We find that the polarization direction of injected spins in each valley is
parallel or anti-parallel to the $z$ direction. The DSP is defined as
\begin{equation}
\mathtt{DSP} = \frac{\zeta}{\hbar\xi/2},\quad \mathtt{DSP^{(I)}_{\{\cdots\}}}=\frac{\zeta^{(I)}_{\{\cdots\}}}{\hbar\xi^{(I)}_{\{\cdots\}}/2}\ .
\end{equation}
Here $\{\cdots\}$ indicates the subscripts for $\mathtt{DSP}^{(I)}$ are the same
as that for $\xi^{(I)}$ and $\zeta^{(I)}$.

\subsection{Carrier injection}
\subsubsection{Band edge carrier injection at 4~K}
\begin{figure}[htp]
  \centering
  \includegraphics[height=5cm]{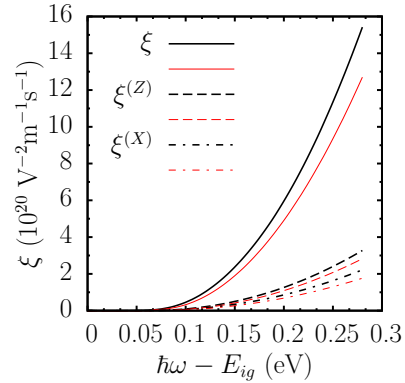}
  \caption{(color online). Spectra of carrier injection rates $\xi$
    (solid curves), $\xi^{(Z)}$ (dashed curves), and  $\xi^{(X)}$
    (dot-dashed curves)  at 4~K. The black thick and red thin curves are the
    results calculated from Eq.~(\ref{eq:X2}) and
    Eq.~(\ref{eq:selectionrule}), respectively. }
  \label{fig:bandedge-total}
\end{figure}
\begin{figure}[htp]
  \centering
  \includegraphics[height=7cm]{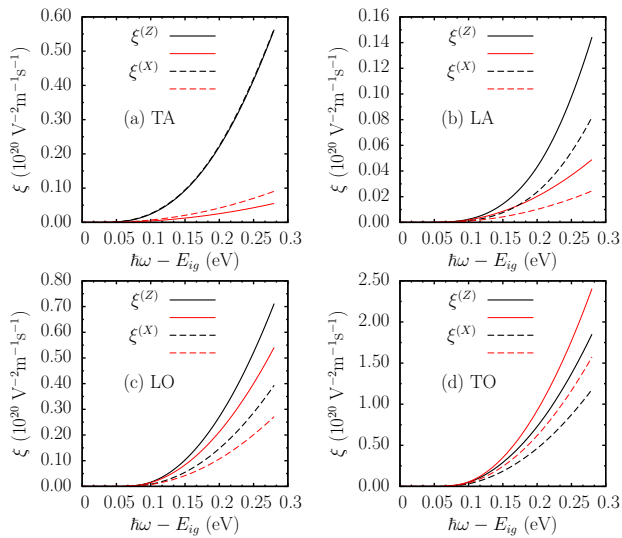}
  \caption{(color online). Phonon-resolved injection rates
    $\xi^{(Z)}_{\tau}$ (solid curves) and
    $\xi^{(X)}_{\tau}$ (dashed curves). The black thick and red thin curves are the
    results calculated from Eq.~(\ref{eq:X2}) and
    Eq.~(\ref{eq:selectionrule}), respectively.}
  \label{fig:bandedge-phonon}
\end{figure}

We first study the carrier injection at the band edge at 4~K. For band
edge injection, only the transitions between the lowest conduction band
and the valence bands need to be considered; the first  
excited conduction band is ignored due to its small density of
states. At this low temperature, only the phonon emission process is important. In
Fig.~\ref{fig:bandedge-total} we show the spectra of the total injection rate
(black thick solid curve)  as well as the injection rates in $Z$
 and $X$ valleys (black thick dashed and dot-dashed curves). 
All injection rates increase with increasing
photon energy. These results are consistent with the analytical results in 
section~\ref{sec:selectionrules}, where the injection rates around
the band edge are approximately proportional to the JDOS. The difference
 between the injection rates in the $Z$ and $X$ valleys shows that the injection is
valley anisotropic. To understand the contribution from each phonon
 branch, we plot phonon-resolved injection rates in
Fig.~\ref{fig:bandedge-phonon}.  We find that each phonon-resolved
spectrum has a shape similar to the total. In our calculation,
the importance of the phonon-assisted processes are in order of TO $>$ LO
$>$ TA $>$ LA, in which the contribution of the LA-assisted process is
less than 5\%. 

At the band edge, the injection rates can also be obtained from the
simplified calculation given by Eq. (\ref{eq:selectionrule}), which can
be used to identify how well the selection rules  in Table
\ref{tab:nz-injection} work. The results of the simplified calculation are also
plotted in Figs.~\ref{fig:bandedge-total} and \ref{fig:bandedge-phonon}
as red thin curves. Compared to the full calculation, Fig.~\ref{fig:bandedge-phonon} shows that these 
results have smaller values for the TA, LA, and LO branches, and a larger
value for the TO branch. The differences are significant for acoustic phonons, and
minor for optical phonons. Even for optical phonons, the
difference is about 30\% at excess photon energy of
$0.25$~eV. However, because the errors for the two most important
phonon branches, {\it i.e.} TO and LO branches, are opposite, the 
difference in the total injection rates between these two methods is
not so great. This difference illustrates not only the
variation of the transition matrix elements on $\bm k_c$ and $\bm k_v$,
but also the failure of the simplified formula at high photon
energy. Nevertheless, the simplified formula gives the correct qualitative 
results.

\begin{figure}[htp]
  \centering
  \includegraphics[height=5cm]{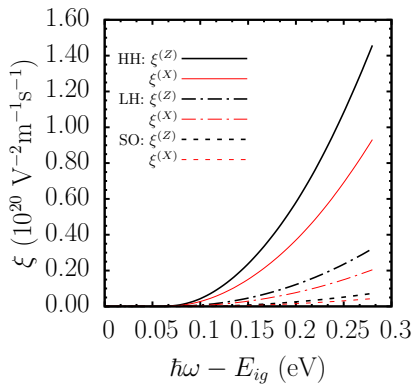}
  \caption{(color online). The contribution from each valence band to the injection rates $\xi^{(Z)}$
    (thick curves) and $\xi^{(X)}$ (thin colored curves) for the TO assisted process.} 
  \label{fig:detailcarrier}
\end{figure}

To better understand the details of carrier injection, we also plot
the contribution from each valence band to $\xi^{(X,Z)}$ for the TO-assisted
process in Fig.~\ref{fig:detailcarrier}. There are two important
features: First, all injection rates increase with photon 
energy, which can be understood by the increase of the involved JDOS
with photon energy. Second, the HH band gives the greatest
contribution to the injection rate, and the SO band gives the
smallest. As the transition matrix elements are the same for these 
three bands at the band edge, the 
magnitude is determined only by the density of states of each valence
band. Similar results are obtained for the other three phonon branches.

\subsubsection{Comparison with experiment at 4~K}
\begin{figure}[htp]
  \centering
  \includegraphics[height=6cm]{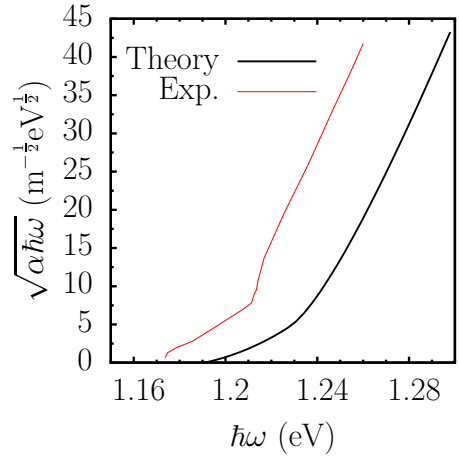}
  \caption{Spectra of $\sqrt{\alpha
      \hbar\omega}$. The solid curve is our theoretical result. 
    The red thin curve is experimental data\cite{Phys.Rev._111_1245_1958_Macfarlane}.}
\label{fig:bandedgecompare}
\end{figure}
Fig.~\ref{fig:bandedgecompare} gives the photon energy dependence of $\sqrt{\alpha(\omega)
  \hbar\omega}$ at $T=4$~K. Here $\alpha(\omega)=\hbar\omega\xi(\omega)/(2n(\omega)c\epsilon_0)$
is the absorption coefficient, $n(\omega)$ is the refractive index of
silicon, $c$ is the speed of light, and $\epsilon_0$ is the vacuum
permittivity. Our result (the solid curve) shows two important
features of the indirect gap injection: 
Near the onset of absorption, for $\hbar\omega<E_{ig}+\hbar\Omega_{\bm
  k_c^0,\mathtt{LO}}^0$, the TA phonon emission process dominates,
then the optical phonons make important contributions for
$\hbar\omega>E_{ig}+\hbar\omega_{\bm k_c^0,\mathtt{LO}}$. The
separation between these two regions is indicated by the kink in the figure. In each region, 
the lineshape of $\sqrt{\alpha \hbar\omega}$ is approximately a linear
function of $\hbar\omega$, which is consistent with the results 
in the parabolic band approximation, {\it i.e.}, $\sqrt{\alpha\hbar\omega}\propto \hbar\omega - E_{ig} -
\hbar\Omega_{\bm
  k_c^0\lambda}$\cite{Phys.Rev._127_765_1962_Hartman,Phys.Rev._108_1384_1957_Elliott}. 
With respect to these features, our results match the experimental results very well.

Yet compared to the experiment by Macfarlane {\it et
  al.} \cite{Phys.Rev._111_1245_1958_Macfarlane}, our result
fails to show the correct energy shift and lineshape at the
beginning of each region. Both of these features are related to the
excitonic effect, which is absent in our calculation. When the excitonic 
effect is considered, both the bound and continuum exciton states
contribute to the absorption, as discussed in detail by Elliott:
\cite{Phys.Rev._108_1384_1957_Elliott} The onset of absorption is
shifted to lower energy by the presence of exciton bound states, and
modified to give a line shape $\alpha\propto\sqrt{\hbar\omega-(E_{ig}+\hbar\Omega_{\bm
    k_c^0\lambda}-E_{ex})}$ associated with the first exciton bound
state and each phonon branch, with an exciton bind energy $E_{ex}\approx
- 14$~meV;\cite{J.Appl.Phys._45_1846_1974_Bludau} the lineshape of the
absorption associated with the exciton continuum states is similar to
that calculated without including the electron-hole
interaction. Despite all this, in the next we will see
that the absorption at high photon energy can still be
well described by our model. Even at low photon energy, we expect our model can 
give a reasonable description of the DSP, because the excitonic
effects are to good approximation spin independent.

\subsubsection{Carrier injection at high temperature}
\begin{figure}[htp]
  \centering
  \includegraphics[height=5cm]{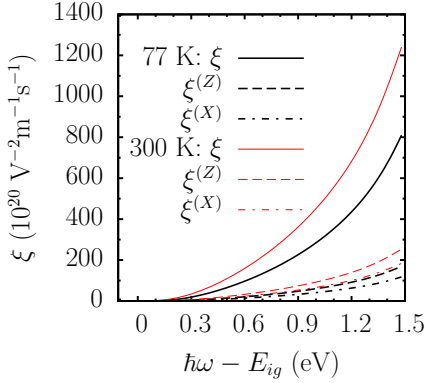}
  \caption{(color online). Spectra of $\xi$, $\xi^{(Z)}$, and $\xi^{(X)}$ at
    77~K (black thick curves) and 300~K (red thin curves).}
  \label{fig:comparison_hightem}
\end{figure}
The indirect gap injection rates depend on the phonon number, and thus
on the temperature. With increasing temperature, the phonon 
absorption processes come into play, and the onset of the injection spectrum moves to lower
photon energy.  The equilibrium phonon numbers $N_{\bm k_c^0\lambda}$ at
different temperatures are as follows: At 77~K, all phonon
numbers can approximately be ignored, so only the emission processes
occur. At 300~K, the phonon numbers become $0.92$ for TA, 
$0.23$ for LA, $0.15$ for LO, and $0.12$ for TO. Because the TA phonon
has smaller energy, the absorption induced 
by the TA phonon-assisted process is the most temperature-sensitive.

\begin{figure}[htp]
  \centering
  \includegraphics[height=6cm]{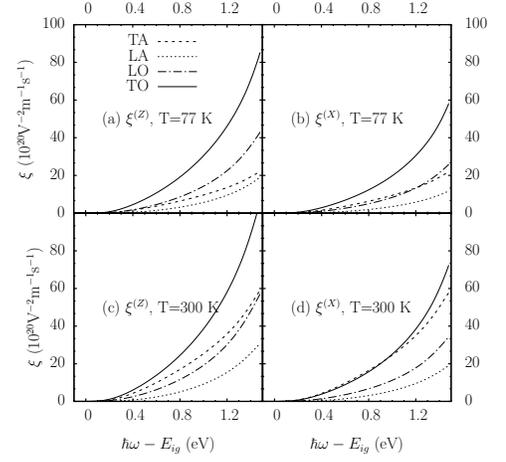}
  \caption{Phonon-resolved $\xi^{(Z)}$ and $\xi^{(X)}$ at $77$~K and $300$~K.}
  \label{fig:carrier_high}
\end{figure}

In Fig.~\ref{fig:comparison_hightem} the spectra of
$\xi$, $\xi^{(Z)}$ and $\xi^{(X)}$ are plotted for 77~K (black
thick curves) and 300~K (red thin curves). All injection rates
increase remarkably with temperature: $\xi$ displays about 50\% increase
when the excess photon energy is 1.5~eV.  Note that
the injection rates at 77~K and 4~K are approximately the same. The
phonon-resolved absorption spectra at different temperatures 
are plotted in Fig.~\ref{fig:carrier_high}. The injection rate from
each phonon branch increases with temperature, and the increment is
most significant for the TA phonon branch. At 300~K, $\xi^{(X)}_{TA}$ is
comparable to $\xi^{(X)}_{TO}$; at low temperature, $\xi^{(X)}_{TA}$
is less than half of $\xi^{(X)}_{TO}$.

\subsubsection{Comparison with experiment at high temperature}
\label{eq:cmph}
\begin{figure}[htp]
  \centering
  \includegraphics[height=5cm]{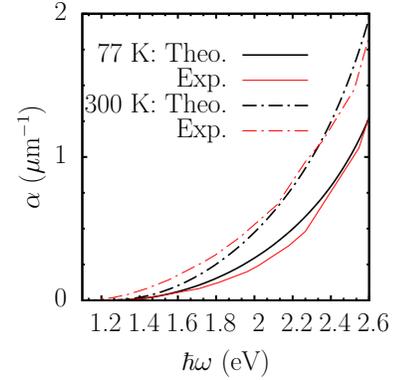}
  \caption{(color online). The comparison of $\alpha$ between our theory (thick curves) and
    experiments\cite{Sze} (thin colored curves) at $77$~K
    (solid curves) and $300$~K (dash-dotted curves).}
  \label{fig:comparison_high}
\end{figure}
In Fig.~\ref{fig:comparison_high}, we compare our results for the
absorption coefficient $\alpha$ with experimental results\cite{Sze} at
$77$ and $300$~K. Compared to the results near the band edge, the
calculations without including excitonic effects fit 
the experiments better at high photon energy. The difference between
theory and experiment at low photon energy can be attributed to
excitonic effects and the temperature dependence of the indirect band gap
energy, neglected in our calculation which takes that energy as its
zero temperature value. Clearly our neglect of excitonic effects and of
this gap energy shift has less consequence at higher energies. In the calculation,
we use the experimental frequency-dependent refractive index
$n(\omega)$\cite{Prog.Photovolt:Res.Appl._3_189_1995_Green}. Because
of the increase in phonon number, the absorption coefficient at 300~K is
remarkably larger than at 77~K. 

\subsubsection{Temperature dependence of carrier injection} 

\begin{figure}[htp]
  \centering
  \includegraphics[height=5cm]{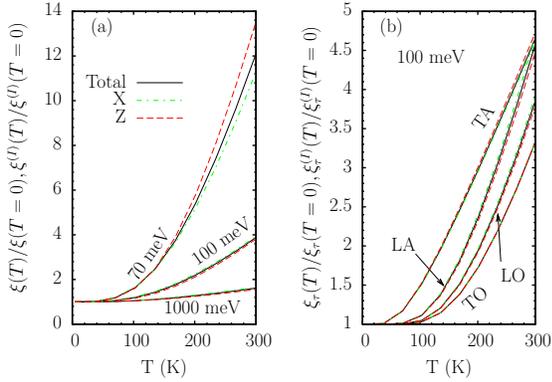}
  \caption{(color online). (a) Temperature dependence of
    $\xi(T)/\xi(T=0)$, $\xi^{(X)}(T)/\xi^{(X)}(0)$, and
    $\xi^{(Z)}(T)/\xi^{(Z)}(0)$ for excess photon energy at
    70, 100, and 1000~meV. (b) Temperature dependence of phonon resolved ratio for
    excess photon energy at 100~meV.  Black solid curves:  
    the total injected carriers, green dash-dotted curves: values in
    the $X$ valley, red dashed curves: values in the $Z$ valley.} 
  \label{fig:carrier_temp}
\end{figure}
Fig.~\ref{fig:carrier_temp} (a) gives the temperature dependence of
the ratio $\xi(T)/\xi(0)$ (black solid curves), $\xi^{(X)}(T)/\xi^{(X)}(0)$ (green
dash-dotted curves), and $\xi^{(Z)}(T)/\xi^{(Z)}(0)$ (red
dashed curves) for $\hbar\omega-E_{ig}=$  70, 100, and 1000~meV. The injection rates increase with
temperature for all photon energies, and their slopes decrease with photon energy. When the
temperature is lower than 70~K, the injection rates are almost independent
of temperature, while they become nearly a linear function of temperature at
temperature higher than 200~K. Their slopes depend on the photon
energy with larger slopes at lower photon energies. The slopes for the
injection in the $Z$
and $X$ valleys are found to be different at low photon
energy, with the difference tending to disappear at high photon
energy. The temperature dependence of phonon-resolved injection rates for excess
photon energy $100$~meV are plotted in Fig.~\ref{fig:carrier_temp} (b). The temperature 
dependence of the phonon-resolved injection rates are similar to the
total, but show different slopes for different branches.  In order to give an
comprehensive understanding of these results, we use the simplified
formula for the band edge injection in
Eq. (\ref{eq:selectionrule}). At high temperature, the phonon number
$N_{\bm k_c^0\lambda}$ is approximately $k_BT/\hbar\Omega_{\bm k_c^0\lambda}$, and
Eq. (\ref{eq:selectionrule}) for absorption near the band edge can be rewritten as  
\begin{equation}
  \frac{\xi^{(I)}_{cv\lambda}(T, \omega)}{\xi^{(I)}_{cv\lambda}(0,
  \omega) } = 1 +
    \frac{k_BT}{\hbar\Omega_{\bm  k_c^0\lambda}}\left[1+\left(\frac{\hbar\omega-E_{ig}+\hbar\Omega_{\bm
            k_c^0\lambda}}{\hbar\omega-E_{ig}-\hbar\Omega_{\bm
            k_c^0\lambda}}\right)^2\right]\ .
\end{equation}
The linear dependence at high temperature is obvious. The
slope is determined by the photon energy, and decreases as the photon
energy increases. At photon energy displayed in
Fig.~\ref{fig:carrier_temp} (b), however, the dependence of the slope
on phonon energy is too complicated to be described by this simple formula.

\subsection{Spin injection}
\subsubsection{Band edge spin injection at 4~K}
\begin{figure}[htp]
  \centering
  \includegraphics[height=6cm]{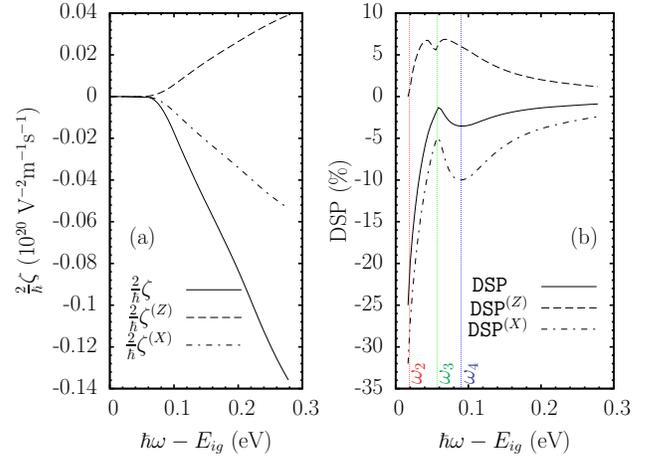}
  \caption{(color online). (a) Spectra of spin injection rates $\zeta$ (solid curves),
    $\zeta^{(Z)}$ (dashed curves), and $\zeta^{(X)}$ (dash-dotted curves) at
    4~K. (b) The corresponding DSP spectra. The labelled
    energies are $\hbar\omega_{2,3,4}-E_{ig}=19,57,90$~meV, respectively.} 
  \label{fig:spin_total}
\end{figure}
\begin{figure}[htp]
  \centering
  \includegraphics[height=6cm]{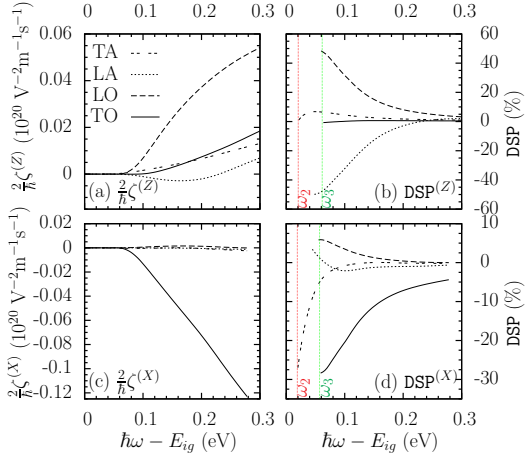}
  \caption{(color online). Phonon-resolved spin injection rates [(a) and (c)] and DSP
    [(b) and (d)] in the $Z$ [(a) and (b)] and $X$ [(c) and (d)] valleys.} 
  \label{fig:spin_phonon}
\end{figure}

In Fig.~\ref{fig:spin_total} (a) we show the spectra of the  spin injection
rate in the $Z$ valley (dashed curve) and
the $X$ valley (dash-dotted curve) as well as the total (solid curve)
at 4~K. The spin injection rates increase with photon 
energy, and show strong valley anisotropy, as do in 
carrier injection.  But the photon energy dependence of the spin
injection rates does not follow the quadratic JDOS, but nearly a linear
function. This indicates the strong wavevector dependence of
$\zeta^{ab}_{c\bm k_cv\bm k_v\lambda}$ given in
Eq.~(\ref{eq:transitionmatrix}). The injected spins in the $X$ and $Z$
valleys have opposite polarization direction. Fine
structures of spin injection can be found from the 
DSP spectra given in Fig.~\ref{fig:spin_total} (b). The total DSP
at the band edge ($\hbar\omega_2-E_{ig}=19$~meV, corresponding to the TA phonon
emission process) is about $-25\%$, while the DSP
in the $Z$ and $X$ valleys approximates $0$ and $-32\%$. When the photon
energy is higher than $\hbar\omega_2$, the spectra can be divided into
three regions: i) $\hbar\omega\in[\hbar\omega_2,\hbar\omega_3]$ with
$\hbar\omega_3-E_{ig}=57$~meV ($=\Omega_{\bm
  k_c^0,\mathtt{TO}}$). Here the $\mathtt{DSP}$ and the
$\mathtt{DSP}^{(X)}$ decrease rapidly with increasing energy to a minimum
value of $-1\%$ and $-5\%$, respectively, while the $\mathtt{DSP}^{(Z)}$
increases to a maximum value of $5\%$ when the photon energy is slightly smaller than $\hbar\omega_3$, and
then decreases slightly to a local minimum at $\hbar\omega_3$. ii)
$\hbar\omega\in[\hbar\omega_3,\hbar\omega_4]$ with
$\hbar\omega_4-E_{ig}=90$~meV. Here the $\mathtt{DSP}$ and the 
$\mathtt{DSP}^{(X)}$ increase to a maximum value $-4\%$ and $-10\%$,
and the $\mathtt{DSP}^{(Z)}$ first slightly increases and then decreases. iii)
$\hbar\omega>\hbar\omega_4$. All $\mathtt{DSP}$ decrease
monotonically to zero. 
The special photon energy $\hbar\omega_3$ is related to the TO
phonon energy given in Sec.~\ref{sec:epm}, the features here are
formed by the contributions from different phonon branches, which are
plotted in Fig.~\ref{fig:spin_phonon}.  

According to the phonon energies, the injection edge energies of different phonon
branches are ordered as TA $<$ LA $<$ LO $<$TO, which is obviously shown in
Figs.~\ref{fig:spin_phonon} (b) and (d). For the spin injection in the
$Z$ valley given in Fig.~\ref{fig:spin_phonon} (a), the contributions from
all phonon branches are almost of the same order of magnitude, even including
the LA phonon branch, which gives a negligible contribution to carrier injection. 
Among these branches, the LO branch gives the largest contribution. 
The corresponding DSP are shown in Fig.~\ref{fig:spin_phonon} (b). 
The DSP induced by TA branches starts from zero, reaches a maximum
at about 45~meV above the onset of absorption, and then
decreases, while the LA phonon branch contributes a negative spin
  injection rate. LO and TO phonons take
effect for $\hbar\omega>\hbar\omega_3$. Therefore, for the total DSP 
in the $Z$ valley (in Fig.~\ref{fig:spin_total}), the first increase
to the maximum is induced by TA phonons, the following dip is
induced by LA phonons, and the second peak is induced by LO
phonons. Figs.~\ref{fig:spin_phonon} (c) and (d) give the spin 
injection rates and DSP in the $X$ valley. Here the contribution from
the TO phonon branch dominates the spin injection, and
those from other phonon branches only give minor contributions. As in
the $Z$ valley, the spin injection in the $X$ valley (in
Fig.~\ref{fig:spin_total}) starts from the TA
phonon branch; its DSP decreases with photon
energy from a nonzero band edge value, and gives the fast decrease
for photon energy in $\hbar\omega\in[\hbar\omega_2,\hbar\omega_3]$. 
At $\hbar\omega_3$ the TO phonons come into play, leading to injected
electrons with a large spin polarization. As the contribution to the
total spin polarization from these electrons begins to dominate, the
DSP reaches a maximum at $\hbar\omega_4$, and decreases at high
frequencies.

\begin{figure}[htp]
  \centering
  \includegraphics[height=6cm]{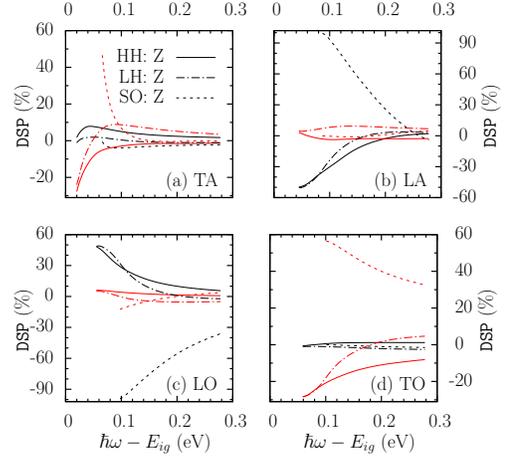}
  \caption{(color online). Phonon and valence band resolved DSP in the
    $Z$ valley (thick
    curves) and $X$ valley (thin colored curves).} 
  \label{fig:spinp}
\end{figure}
We now turn to resolving the spin injection rates into contributions
from different valence bands. The results are similar to the
corresponding resolution of carrier injection rates. The HH band gives the largest contribution due to
its large density of states, while the SO band gives the
smallest. However, there are subtleties in the DSP of injected spins
from each valence band. In Fig.~\ref{fig:spinp} we give the valence band and
phonon-resolved spectra of the DSP. While the band edge values
of the DSP are consistent with results in Table\ \ref{tab:bevalue2}, the
spin polarization direction shows a complicated energy dependence
based on whether or not the band edge DSP is zero. For
processes with nonzero band edge values, the DSP decrease from the nonzero
values. For processes with zero band edge values, the DSP increase to
a maximum value, and then decrease. Some of these DSP change sign with
increasing photon energy.

\subsubsection{Spin injection at high temperature}
\begin{figure}[htp]
  \centering
  \includegraphics[height=6cm]{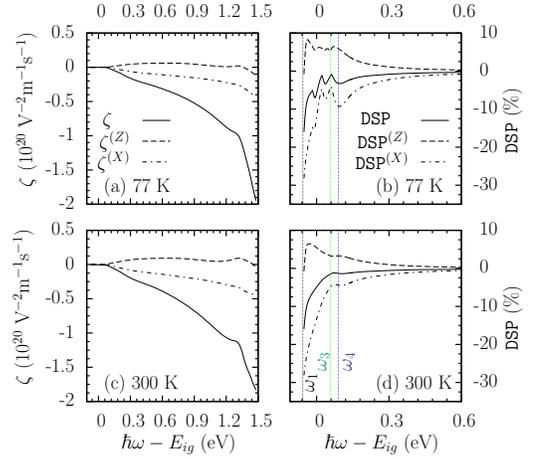}
  \caption{(color online). Spectra of spin injection rates   and DSP
    at 77~K  and 300~K. $\hbar\omega_1-E_{ig}=-57$~meV.}
  \label{fig:spin_high}
\end{figure}
\begin{figure}[htp]
  \centering
  \includegraphics[height=6cm]{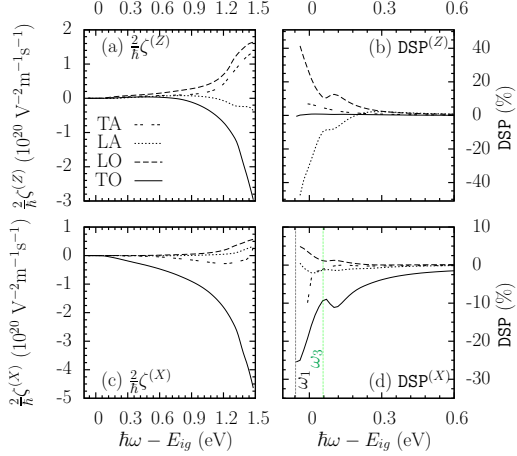}
  \caption{(color online). Phonon-resolved spectra of spin injection rates and DSP at
    300~K.}
  \label{fig:spin_high_phonon}
\end{figure}

In Fig.~\ref{fig:spin_high} we plot the
spectra of spin injection rates at 77~K and 300~K. In contrast to the
carrier injection rates, which increase considerably from 77~K to
300~K, both the total spin injection rates and the rates in each valley
change little from 77~K [Fig.~\ref{fig:spin_high} (a)] to 300~K
[Fig.~\ref{fig:spin_high} (c)]. In order to show clearly the spin injection at
the injection edge, the DSP are plotted in (b) and (d)
respectively. the phonon absorption
process becomes increasingly important as the temperature rises, and is shown by the left shifts of the
onset of DSP to $\hbar\omega_1$, which is determined by the TO/LO phonon
absorption processes. In addition, the
peak appearing at $\hbar\omega_4$ and 4~K in
Fig.~\ref{fig:spin_total}~(b) becomes fairly obscure at high
temperature. These results can be better
understood from the phonon-resolved  DSP spectra at 300~K, given in
Fig.~\ref{fig:spin_high_phonon}.

\begin{figure}[htp]
  \centering
  \includegraphics[height=6cm]{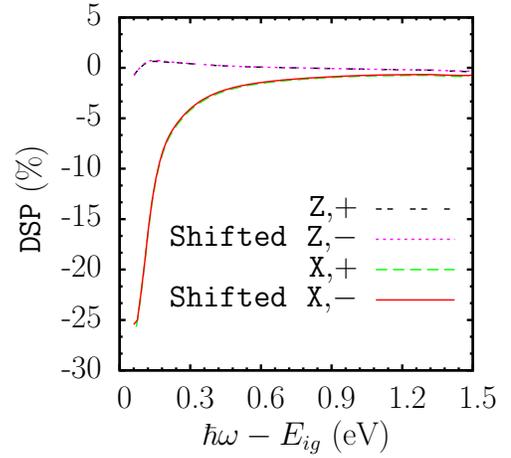}
  \caption{(color online). Spectra of DSP for TO phonon absorption and
    emission processes in  the $Z$ and
    $X$ valleys at 300~K.}
  \label{fig:spin_phonon_twoprocess}
\end{figure}

For the spin injection 
rates in the $Z$ valley, shown in Fig.~\ref{fig:spin_high_phonon} (a), the
contribution from each phonon branch is still of the same order of
magnitude, but the TO phonon-assisted process dominates the injection 
at high photon energy. For the spin injection rates in the $X$ valley, shown
in Fig.~\ref{fig:spin_high_phonon} (c), the contribution from the TO
phonon branch dominates for all photon energies. However, the DSP,
plotted in Figs.~\ref{fig:spin_high_phonon} (b) and (d), show finer structure in the low photon energy region than the DSP at
4~K. The fine structure is
induced by the combined effect of the phonon emission and
phonon absorption processes, which take effect at 300~K with
considerable equilibrium phonon numbers for all phonon branches.
According to the simplified Eq. (\ref{eq:selectionrule}), the
injection rates for phonon absorption/emission processes differ
from the JDOS, $J_{cv\lambda}(\hbar\omega\mp\hbar\Omega_{\bm
  k_c^0\lambda})$, at low photon energy. Therefore, spectra of DSP for
these two processes have similar lineshapes, except 
 for a phonon energy shift left or right. This conclusion is confirmed by our
numerical results, shown in
Fig.~\ref{fig:spin_phonon_twoprocess}. Here the DSP from TO phonon
absorption and emission processes are plotted for spins in both $Z$ and $X$ valleys at 300~K, and all the
absorption curves are right shifted by $2\Omega_{\bm k_c^0,TO}$. The overlap is
obvious. Based on these results, we return to Fig.~\ref{fig:spin_high_phonon}.

We consider the DSP from the TO phonon-assisted process in the $X$ valley, shown as
solid curve in Fig.~\ref{fig:spin_high_phonon} (d), is taken as an example. The
spectrum in $[\hbar\omega_1,\hbar\omega_{3}]$ is induced by the phonon
absorption processes, and decreases with increasing photon energy. When the photon
energy is higher than $\hbar\omega_3$, the TO phonon emission process
comes into play. However,
the spin injection rate for phonon emission is approximately
proportional to $1+N_{\bm k_c^0,TO}$, while that for phonon absorption
is proportional  to $N_{\bm k_c^0,TO}$; thus the former gives
much larger injection rates than the latter. This results in a peak in
the total DSP for the TO phonon-assisted process, at energies where phonon emission
becomes important.

\subsubsection{Temperature dependence of spin injection}
\begin{figure}[htp]
  \centering
  \includegraphics[height=6cm]{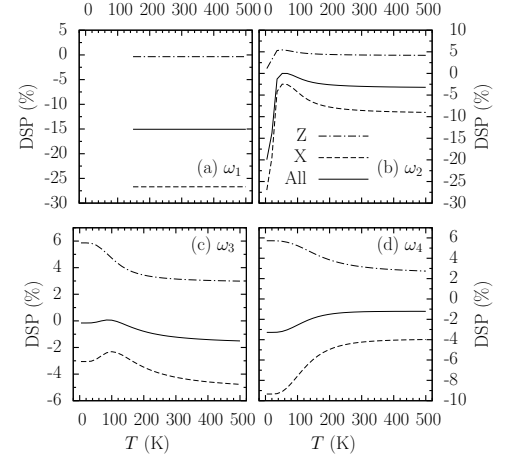}
  \caption{Temperature dependence of $\mathtt{DSP}^{(Z)}$ (dot-dashed
    curves), $\mathtt{DSP}^{(X)}$ (dashed curves), as well as the
    total DSP (solid curves) at photon energies 
    $\hbar\omega_{1-4}$.}
  \label{fig:temperature}
\end{figure}
\begin{figure}[htp]
  \centering
  \includegraphics[height=6cm]{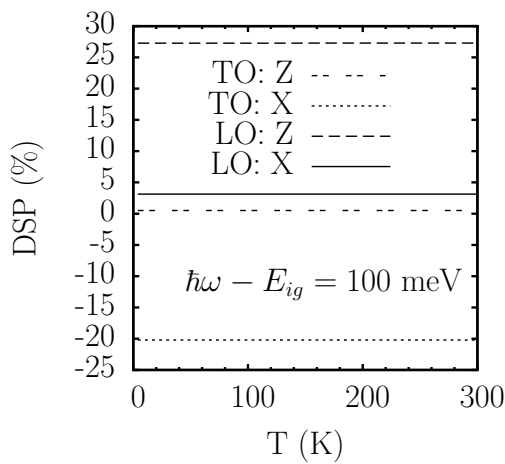}
  \caption{Temperature dependence of the DSP for TO and LO phonon emission
    assisted processes.}
  \label{fig:temperature_branch}
\end{figure}

Fig.~\ref{fig:temperature} shows the temperature dependence of the DSP
at the different photon energies $\hbar\omega_{1-4}$. 
For photon energy $\hbar\omega_1$, 
all DSPs are temperature-independent: The TO phonon
absorption process is the only process at this photon energy, and the injection rates of carriers and spins
are approximately proportional to $N_{\bm k_c^0,TO}$; the two rates
therefore scale the same with temperature, leading to a
temperature-independent DSP. We check the temperature dependence of the DSP for
all phonon branches at photon energy $\hbar\omega-E_{ig}=100$~eV
in Fig.~\ref{fig:temperature_branch}. All these DSP are independent of 
temperature. This conclusion is valid for all photon
energies considered to this point in this paper. However, the
total DSP, shown in Figs.~\ref{fig:temperature} (b), (c), and (d),
{\it are} temperature
dependent.  At the indicated photon energies, more than one phonon
branch makes a contribution. The injection rates for different phonon
branches have different temperature
dependences due to the different phonon energies. The ratio between
the total injection rates of carriers and spins, {\it i.e.} the DSP,
thus becomes temperature-dependent. 
At high temperature, $N_{\bm q\lambda}\propto T$ is approximately
valid for all phonon branches, so the injection coefficients of $\xi$
and $\zeta$ are approximately proportional to $T$ and
give a constant DSP as saturation value. This is shown in Fig.~\ref{fig:temperature}.

\section{conclusion}
In conclusion, we have performed a full band structure calculation to
investigate the indirect optical carrier and spin injection in bulk
silicon. The injection spectra for carriers, spins, and DSP in each
valley are studied in detail at 
different temperatures. The indirect gap injection is dominated by the
transition from the HH band
to the lowest conduction band  due to the large JDOS. When 
incident light is propagating along one principal axis, the injection shows strong valley anisotropy. The injection
rates induced by each phonon-assisted process increase with
temperature. For carrier injection, we find that in the $Z$ valley the TO phonon-assisted
process dominates up to 300~K; in the $X$ valley, it only
dominates at low temperature, while the injection rates induced by the
TA phonon-assisted process increases to a comparable value at
300~K. The higher the photon energy, the weaker the temperature
dependence. For spin injection, we find that injected spins in the 
$Z$ and $X$ valleys have opposite polarization directions. In the $Z$ valley, the LO phonon-assisted process dominates
around the band edge, while the TO phonon-assisted process dominates at high photon energy; in the $X$
valley, the TO phonon-assisted process dominates for all photon energies. 

The calculated absorption coefficients are in
good agreement with experiments at high photon energy. Experience from
direct gap absorption might lead one to believe this would not hold,
since first principle studies of direct optical absorption in Si and 
GaAs\cite{Phys.Rev.B_62_4927_2000_Rohlfing,Rev.Mod.Phys._74_601_2002_Onida,Phys.Rev.B_71_195209_2005_Leitsmann}
demonstrated that the electron-hole interaction plays an
important role even for high photon energy. Full band structure
calculations of indirect gap absorption, including electron-hole
interaction, have yet to be done, and the effect of these interaction
on the indirect absorption coefficient at high photon energies have yet to be
established. In any event, since to first approximation the nearly
spin-independent electron-hole interaction will modify spin and
carrier injections in the same way, it is reasonable to expect that the DSP of
  injected electrons in exciton continuum states will
be insensitive to this interaction, which we neglect. Future work to
confirm this is clearly in order, but in the interim we feel our study
constitutes a good first investigation.

The DSP spectra excibit a rich variety of behaviours. At 4~K, the maximum
DSP is about $-25\%$ for photon energy $E_{ig}+\Omega_{\bm
  k_c^0,TA}$; absorption at this temperature and energy is dominant by
TA phonon emission process. 
With increasing temperature, the phonon absorption processes
become important, and the DSP for this photon energy decreases quickly. At
300~K, the maximum DSP appears at the photon energy
$E_{ig}-\Omega_{\bm k_c^0,TO}$ as a value $-15\%$, which only comes 
from the TO phonon absorption assisted process, and is
temperature-independent. The DSP in the $X$ valley can reach a
maximum value $-32\%$ at 4~K and $-26\%$ at 300~K; both are larger than the total
value and the value in the $Z$ valley. Compared to bulk silicon, it
should therefore be more efficient
to inject spin in a confined silicon structure, where conduction
band valleys are splitted.

\begin{acknowledgements}

This work was supported by Natural Sciences and Engineering Research
Council of Canada. J.L.C acknowledges support from China Postdoctoral
Science Foundation. J.R. acknowledges support from
FQRNT. J.F. acknowledges support from DFG SPP1285.
\end{acknowledgements}

\appendix
\section{Modified Tetrahedron Summation}
\label{app:numerical}
Both Eq.~(\ref{eq:X2}) and the JDOS in Eq.~(\ref{eq:jointdos}) are of
the following form
\begin{eqnarray}
  X(\omega) &=& \int_{V_c^{\prime}}d\bm k_c\int_{V_v^{\prime}}d\bm
  k_v\delta(\varepsilon_{\bm k_c}-\varepsilon_{\bm
    k_v}+\hbar\Omega_{\bm k_c-\bm k_v}-\hbar\omega)\nonumber\\
&&\times X(\bm
  k_c, \bm k_v)\ .
\label{eq:app1}
\end{eqnarray}
Here the integration spaces $V_c^{\prime}$ and $V_v^{\prime}$ are the
irreducible BZ wedges, which are confined by the $\Gamma$, $X$, $L$, $U
(K)$, and $W$ points of a fcc Brillouin zone. Our numerical scheme is
based on the improved LATM \cite{Phys.Rev.B_49_16223_1994_Blochl,Phys.Rev.B_76_205113_2007_Nastos}. 
We divide $V_c^{\prime}$ ($V_v^{\prime}$) into $N_c$ ($N_v$) small
tetrahedra as $V_c^{\prime}=\sum_{i=1}^{N_c}V_{c,i}^{\prime}$
($V_v^{\prime}=\sum_{j=1}^{N_v}V_{c,j}^{\prime}$) with vertices $\bm
k_{c,I}, I=1,\cdots,M_c$ ($\bm k_{v, J}, J=1,\cdots,M_v$). 
By a simple transformation, Eq.~(\ref{eq:app1}) becomes
\begin{eqnarray}
  X(\omega) &=& \sum_{ij}\int d\epsilon X_{ij}(\omega,\epsilon)\
  ,\nonumber
\end{eqnarray}
with 
\begin{eqnarray}
X_{ij}(\omega,\epsilon) &=& \int_{V_{c,i}^{\prime}}d\bm k_c
  \delta(\varepsilon_{\bm k_c} - \epsilon) Y_j(\bm k_c, \hbar\omega -\hbar\bar{\Omega}_{ij}-
  \epsilon)\ , \nonumber\\\label{eq:app2}
\end{eqnarray}
and
\begin{eqnarray}
Y_j(\bm k_c,\epsilon) &=& \int_{V_{v,j}^{\prime}}d\bm
  k_v\delta(\varepsilon_{k_v}-\epsilon)X(\bm
  k_c, \bm k_v)\ .
\label{eq:app3}
\end{eqnarray}
In obtaining the equations above, we approximate the phonon energy to
be a fixed value $\hbar\bar{\Omega}_{ij}$ between the two small
tetrahedra $V_{c,i}^{\prime}$ and $V_{v,j}^{\prime}$. Now the
Eqs.~(\ref{eq:app2}) and (\ref{eq:app3}) have the same 
shape, so we only discuss Eq. (\ref{eq:app3}) in the
following. Though the tetrahedron method can be applied to
Eq. (\ref{eq:app3}) directly, we avoid doing this because the transition
matrix elements are obtained within the scheme of the EPM, which are
very time-consuming in calculation for each $(\bm k_{c,I}, \bm k_{v, J})$ pair. However, the direct tetrahedron
method needs more $\bm k$ points for convergence than is
computationally feasible. Instead,
since the dependence of the transition matrix elements on $\bm k_c$
and $\bm k_v$ is weak, we linearly interpolate $X(\bm k_c,\bm k_v)$ in tetrahedron 
$V_{v,j}^{\prime}$ as  
\begin{equation}
X(\bm k_c, \bm k_v) = \sum_{m=1}^4X(\bm k_c, \bm k_{v, j}^m)
F_{v,j}^m(\bm k_v)\ ,\nonumber
\end{equation}
with $\bm k_{v,j}^m$ being the position of four vertices of this tetrahedron, and
$F_{v,j}^m(\bm k_v)$ being the linear interpolation function related
to the $m^{th}$ vertex. Then Eq.~(\ref{eq:app3}) becomes
\begin{eqnarray}
  Y_j(\bm k_c, \epsilon) &=& \sum_{m=1}^4w_{v,j}^m(\epsilon)X(\bm
  k_c,\bm k_{v,j}^m)\ ,\nonumber\\
  w_{v,j}^m(\epsilon) &=& \int_{V_{v,j}^{\prime}} d\bm k_v
  \delta(\varepsilon_{v\bm k_v}-\epsilon) F_{v,j}^m(\bm k_v)\ .
\end{eqnarray}
In the same way, we get
$w_{c,i}^n(\epsilon)$ for Eq. (\ref{eq:app2}).  Then we find
\begin{eqnarray*}
X_{ij}(\omega,\epsilon) &=&  \sum_{nm=1}^4w_{c,i}^n(\epsilon)w_{v,j}^m(\hbar\omega-\hbar\Omega_{\bm
    k_{c,i}^n-\bm k_{v,j}^m}-\epsilon)\nonumber\\
&&\times X(\bm k_{c,i}^n,\bm k_{v,j}^m)\ ,
\end{eqnarray*}
and
\begin{eqnarray}
  X(\omega) &=& \int d\epsilon
  \sum_{IJ}W_{c,I}(\epsilon)W_{v,J}(\hbar\omega-\hbar\Omega_{\bm
    k_{c,I}-\bm k_{v, J}}-\epsilon)\nonumber\\
&&\times X(\bm k_{c,I}, \bm k_{v, J})\ ,
\label{eq:final}
\end{eqnarray}
with 
\begin{eqnarray}
  W_{c,I}(\epsilon)&=&\sum_{in,
  k_{c,i}^n=k_{c,I}}w_{c,i}^n(\epsilon)\ ,\nonumber\\
W_{v,J}(\epsilon)&=&\sum_{jm,
k_{v,j}^n=k_{v,J}}w_{v,j}^n(\epsilon)\ .
\end{eqnarray}
In the above equations, we replace $\hbar\Omega_{\bm k_{c,i}^n-\bm
  k_{v,j}^m}$ by $\hbar\bar{\Omega}_{ij}$.

In indirect absorption, the excited holes are located
around the top of the valence band, while the electrons are around the
band edge of the conduction band. So we use a different division for $\bm
k_c$ and $\bm k_v$. With the division points $\bm k_{c,I}$ and $\bm
k_{v,J}$, we use the Delaunay triangulation method from the CGAL
package\cite{cgal} to set up the tetrahedra. In these tetrahedra,
the weights $w_{v,j}^m$ and $w_{c,i}^n$ can be calculated. However, the
integration includes a $\delta$ function, which is a fast-varying
function in $\bm k$ space, and the present tetrahedron is too rough to
obtain the weights with required precision. We refine this
tetrahedron into smaller ones. 

In our calculation, the division depends on the band structure. For
conduction bands and the heavy hole band, $\bm k_{c,I}$ or $\bm k_{v,J}$
are generated with their distance 
not larger than $0.05\times\frac{2\pi}{a}$, while for light hole and
spin split-off bands, the distances become $0.04\times\frac{2\pi}{a}$
and $0.02\times\frac{2\pi}{a}$, respectively. For the heavy hole grid,
the accurate weights can be obtained by refining each tetrahedron as
$8^3$ smaller ones (each edge is refined into 8 parts).  

The convergence of the results is examined for the injection from
heavy hole band to conduction band by changing the division points
distance to $0.025\times\frac{2\pi}{a}$. The difference between the
injection rates from these two divisions is less than 5\%.

\bibliographystyle{prsty}

\end{document}